# History of the Solar Nebula from Meteorite Paleomagnetism

Benjamin P. Weiss[1]*, Xue-Ning Bai[2]*, and Roger R. Fu[3]

[1]Department of Earth, Atmospheric, and Planetary Sciences, Massachusetts Institute of Technology, Cambridge, MA, USA
[2]Institute for Advanced Study and Department of Astronomy, Tsinghua University, Beijing, China
[3]Department of Earth and Planetary Sciences, Harvard University, Cambridge, MA, USA

*Corresponding authors. Email: bpweiss@mit.edu, xbai@tsinghua.edu.cn

**Abstract**
We review recent advances in our understanding of magnetism in the solar nebular and protoplanetary disks (PPDs). We discuss the implications of theory, meteorite measurements, and astronomical observations for planetary formation and nebular evolution. Paleomagnetic measurements indicate the presence of fields of 0.54 ± 0.21 G at ~1 to 3 astronomical units (AU) from the Sun and ≳0.06 G at 3 to 7 AU until >1.22 and >2.51 million years (Ma) after solar system formation, respectively. These intensities are consistent with those predicted to enable typical astronomically-observed protostellar accretion rates of ~$10^{-8}$ $M_\odot$ y$^{-1}$, suggesting that magnetism played a central role in mass and angular momentum transport in PPDs. Paleomagnetic studies also indicate fields <0.006 G and <0.003 G in the inner and outer solar system by 3.94 and 4.89 Ma, respectively, consistent with the nebular gas having dispersed by this time. This is similar to the observed lifetimes of extrasolar protoplanetary disks.

## INTRODUCTION

Newly-formed Sun-like stars are surrounded by planar distributions of circumstellar material known as protoplanetary disks (PPDs) (*1*). PPDs form as consequence of the collapse of molecular cloud cores under angular momentum conservation and are a critical intermediate stage of solar system formation (*2*). They consist of mostly of hydrogen and helium gas, with about ~1% by mass solids (i.e., dust) that serve as the building blocks for planetary bodies. The hundreds of PPDs currently accessible to astronomical observations have typical radii of ~$10^1$-$10^2$ AU and masses ranging from $10^{-4}$ to $10^{-1}$ (median ~$10^{-2}$) solar masses ($M_\odot$) (*2*). About half of all PPDs disperse between ~2 and 4 Ma after collapse of their parent molecular clouds (*3, 4*). Recently, PPDs have been found to possess rich substructures (*5, 6*), which may suggest that they form planets extremely efficiently.

PPDs actively accrete onto their host protostars. Typical accretion rates are ~$10^{-8}$ $M_\odot$ y$^{-1}$ for the bulk of PPD lifetimes (*7*), but can be several orders of magnitude higher at early phases and drop to <$10^{-10}$ $M_\odot$ y$^{-1}$ as disks disperse (*8*). Accretion requires efficient transport of disk angular momentum and we will shortly show that a prominent role for magnetic fields in this process is likely unavoidable. Depending on the mechanisms involved, the expected field strength can be directly estimated for any given accretion rate. Because angular momentum transport further governs the overall gas dynamics (e.g., flow properties such as the level of turbulence) as well as the long-term mass evolution of the disk, it is the central focus of the theory of PPDs.

Measuring the magnetic field strength in PPDs therefore offers a unique opportunity to test disk theory with far-reaching implications for planet formation. Thus far, astronomical observations have yet to unambiguously constrain the strength and morphology of magnetic fields in the planet-forming regions of PPDs (*9*). However, beginning with the formation of calcium aluminum-rich inclusions (CAIs) at ~4,567 Ma ago and lasting several Ma, our own solar system is thought to have transitioned through a PPD phase known as the solar nebula. The present-day solar system preserves a diversity of records of its formation and evolution dating back to this early epoch, largely in the form of meteorites. Recently, laboratory analyses of natural remanent magnetization (NRM) in meteorites have provided time- and spatially-resolved constraints on the magnetic field intensity in the solar nebula (*10*). Because many meteoritic materials are thought to have been magnetized in the solar nebula or shortly after the solar nebula dispersed, they currently provide the best available data on magnetic fields for testing disk theory its relationship with solar system formation.

Here we review recent advances in the theory of disk magnetic fields and the status and implications of astronomical and meteorite records for PPD magnetism. We discuss how meteorite studies have enabled measurements of field strength (paleointensity) over space and time, and how these measurements can help distinguish between various proposed mechanisms for angular momentum transport, formation of chondrules, and the lifetime of the nebula and its implications for giant planet formation.

**THEORY**
**A. Mechanisms Governing Angular Momentum Transport**
Formation of PPDs from the collapse of prestellar and protostellar cores in molecular clouds is a natural consequence of angular momentum conservation. The subsequent evolution of such a disk into the present-day solar system, in which >99% of the mass is within the Sun and >99% of the angular momentum is carried by the planets, required the inward transfer of mass. Angular momentum conservation, in turn, required that the angular momentum originally carried by this mass be transferred to a smaller amount of mass that flows outward.

In cylindrical coordinates ($R$, $\phi$, $z$), angular momentum conservation is given by (*11*):

$$\frac{\partial(2\pi R \Sigma j)}{\partial t} + \frac{\partial(\dot{M}_{\text{acc}} j)}{\partial R} + \frac{\partial}{\partial R}\left[2\pi R^2 \int_{z_{\text{bot}}}^{z_{\text{top}}} dz \left(\overline{\rho \delta v_R \delta v_\phi} - \overline{\frac{B_R B_\phi}{4\pi}}\right)\right]$$
$$+ 2\pi R^2 \left(\overline{\rho v_z v_\phi} - \overline{\frac{B_z B_\phi}{4\pi}}\right)\Big|_{z_{\text{bot}}}^{z_{\text{top}}} = 0 \qquad (1)$$

where we take the disk to be bounded vertically between vertical locations $z_{\text{bot}}$ and $z_{\text{top}}$, $\Sigma$ is the disk surface density, $j = \Omega_k R$ is the specific angular momentum at radius $R$, $\Omega_k$ is the Keplerian angular velocity, $\dot{M}_{\text{acc}}$ is the accretion rate (defined as negative for inward flows), $\boldsymbol{B} = (B_R, B_\phi, B_z)$ is the magnetic field, $\boldsymbol{v} = (v_R, v_\phi, v_z)$ is the local velocity of disk material, the symbol $\delta$ represents deviation from mean flow velocities, and the overbar indicates time and spatial averages. [We use cgs units in the main text of this paper and have listed Système International (SI) versions of the equations in the supplement (note that 1 G = 100 µT)]. The first two terms represent the rate of change of angular momentum of disk material per unit radius and the net flow of angular momentum at radius $R$ associated with accretion flow, respectively. The terms in the two parentheses in the third and fourth terms correspond to the $R\phi$ and $z\phi$ components of the stress tensor $\boldsymbol{\mathcal{T}}$, representing the angular momentum flux transported along the $R$ and $z$ directions, respectively. Angular momentum transport can be of hydrodynamic (first terms in each parenthesis) and/or magnetic (second terms in each parenthesis) in origin. These terms are known as the Reynolds stress and Maxwell stress, respectively. Note that although there is also a contribution from self-gravity, we have ignored this for brevity and because it is likely important only in the earliest phases of disk formation.

Contribution from positive $\mathcal{T}_{R\phi}$ transports angular momentum radially outwards, causing the inner disk region to lose angular momentum and accrete onto the protostar, whereas the outer disk gains angular momentum and expands. It is custom to parameterize the stress tensor as $\mathcal{T}_{R\phi} \equiv \alpha P$ where $P$ is the pressure and $\alpha$ is a dimensionless constant, resulting in the so-called $\alpha$-disk model (*12*). In steady state with $\dot{M}_{\text{acc}}$ constant in radius and assuming an isothermal equation of state, $P = \rho c_s^2$ for isothermal sound speed $c_s$, we see that the accretion rate is $\dot{M}_{\text{acc}} = 2\pi\alpha\Sigma c_s$. For a standard minimum-mass solar nebula (*13*) disk model with surface density $\Sigma = 1700$ g cm$^{-2}(R/\text{AU})^{-3/2}$ and temperature $T = 280$ K $(R/\text{AU})^{-1/2}$, one finds $\dot{M}_{\text{acc}} \sim 8.3 \times 10^{-8} M_\odot \text{y}^{-1}(\alpha/0.01)(R/\text{AU})^{-1/2}$. Thus, a typical PPD accretion rate of $10^{-8}$ $M_\odot$ y$^{-1}$, if solely driven by radial transport, generally requires $\alpha \sim 10^{-3}$-$10^{-2}$.

The effect of $\mathcal{T}_{z\phi}$ transports angular momentum vertically, carried out by a wind. Note that the hydrodynamic part of $\mathcal{T}_{z\phi}$ does not contribute to the net angular momentum loss because it simply describes the angular momentum of materials advected by the wind and thus does not extract extra angular momentum from the disk. Only the magnetic part of $\mathcal{T}_{z\phi}$ extracts additional angular momentum from the disk, which drives the entire disk to accrete. This illustrates that wind-driven accretion must be magnetic in nature.

From the above, we see that hydrodynamic mechanisms can contribute to radial transport of angular momentum. This is generally attributed to waves and turbulence, particularly requiring correlated fluctuations in radial and azimuthal velocities (i.e., a positive $\overline{\rho\delta v_R \delta v_\phi}$ term). On the other hand, mechanisms involving magnetic fields can transport angular momentum both radially and vertically, for which turbulence is not necessary. Below, we briefly review our current understandings from both perspectives, and conclude that magnetic mechanisms likely dominate.

## B. Hydrodynamic turbulence
**1. Overview.** Hydrodynamic angular momentum transfer mechanisms generally rely on the onset and sustenance of turbulence, which is closely related to the hydrodynamic stability of the disk [see recent reviews (*14, 15*)]. When ignoring thermodynamics, the stability of a disk is described by the well-known Rayleigh criterion (*16*), which states that a rotating inviscid system is linearly stable to axisymmetric perturbations as long as $j$ increases radially outward ($\partial j/\partial R > 0$) (*17*) (Fig. 1). This condition holds for Keplerian disks. As such, studies of hydrodynamic disk instabilities typically fall into two categories.

First, consider the subcritical transition to turbulence. It is well known that linearly stable flows may become unstable to developing turbulence at large Reynolds numbers, where the Reynolds number is defined as $Re \equiv LV/\nu$ for system size $L$, characteristic velocity $V$, and fluid kinematic viscosity $\nu$ (*18*). Triggering such a transition requires finite-amplitude fluctuations and is thus termed "subcritical". However, it is not yet clear whether such a transition occurs for Keplerian rotation profiles [see ref. (*19*)]. An extrapolation of numerical simulations conducted up to $Re \sim 10^4$ in the Rayleigh-stable regime towards Keplerian rotation found the critical Reynolds number $Re_c$ for such a transition to be on the order of $10^{10}$ or higher (*20*), which is comparable to the typical Reynolds numbers of PPDs. However, upon transition to turbulence, the Reynolds stress scales as $Re_c^{-1}$ such that the aforementiond high value of $Re_c$ essentially makes angular momentum transport negligible.

Second, in recent years, several hydrodynamic instabilities have been discovered after considering more realistic disk thermal structures. Assuming adiabatic perturbations, the linear stability of a rotating disk subject to axisymmetric perturbations is described by the so-called Solberg-Høiland criteria (*14, 15, 21*), which combine the Rayleigh criterion mentioned above with the Schwarzschild criterion that convection is driven by a positive entropy gradient. The Solberg-Høiland stability criteria are usually well satisfied in PPDs. However, the criteria may not be applicable because PPDs are not necessarily adiabatic: gas is heated mainly by stellar irradiation, while sufficiently rapid cooling brings gas to equilibrium temperatures on timescales much shorter than the orbital timescale such that the gas is locally isothermal. Only in the opposite limit of slow cooling is the adiabatic regime relevant. Real disks are likely in between these regimes, having different cooling times in different regions depending on physical properties like the local disk structure and dust abundance.

**2. Candidate hydrodynamic instabilities.** Three purely hydrodynamic instabilities have recently been considered for PPDs: the vertical shear instability (VSI), the related processes of convective overstability (COS) and subcritical baroclinic instability (SBI), and the zombie vortex instability (ZVI). A detailed discussion for each of the instabilities is beyond the scope of this review and readers are referred to refs. (*14, 15*) for further information. Here, we summarize the main requirements for each instability to operate, their possible existence in PPDs (see Fig. 2), and the likely outcomes.

The VSI draws its free energy from the vertical gradient of azimuthal velocity $v_\phi$. This vertical shear arises because with irradiation heating, the radial pressure gradient in the disk generally varies with height. The VSI requires the gas to be close to locally isothermal (i.e., having a cooling timescale shorter than ~1% of the local orbital time), a condition that is likely satisfied in a radial range of about 10-100 AU (*22*). Simulations show that the VSI leads to vigorous turbulence dominated by vertical motion. It achieves $\alpha$ values up to $10^{-3}$ in the limit of instantaneous cooling (*23*), and $\alpha$ values of several times $10^{-4}$ for more realistic cooling rates (*24*). These fall short of the $10^{-2}$ value required for them to dominate angular momentum transport in outer disks.

The COS derives its free energy from a radial entropy gradient similar to convection (*25, 26*). While radial convection is inhibited by disk rotation, the COS is achieved by allowing thermal relaxation so that epicyclic oscillations are boosted by buoyancy (*27*). The COS is most efficient when the cooling timescale is comparable to the orbital timescale. The non-linear saturation of the COS is the SBI [a process actually discovered earlier (*28, 29*)] which amplifies existing vortices over hundreds of orbital timescales. The vortices launch density waves which transport angular momentum outwards with $\alpha$ values of up to a few times $10^{-3}$ (*29, 30*). The result strongly depends on the outward radial entropy gradient, which in typical models within 20 AU is likely present mainly in the disk upper layer (rather than the midplane region where it would be more effective) (*15*).

The ZVI derives from the finding that vortices in stably stratified disks can excite "baroclinic critical layers" at some distance via buoyancy waves, which subsequently create copies of themselves (*31, 32*) that eventually lead to turbulence with volume-filling vortices. The ZVI is a nonlinear instability which requires a finite-amplitude vorticity field to be triggered. It also requires the gas to behave nearly adiabatically (i.e., have a relaxation time $\gtrsim$ 10 orbital times), such that the regions in which it may operate are primarily limited to the very optically-thick inner ~1 AU of PPDs (*33*). Numerical studies so far are limited to local simulations, the contribution to angular momentum transport is likely small, and the associated Reynolds stress is well below the desired level (*34*).

In addition to the above processes, there is also the well-known gravitational instability (GI) (*35*). The GI is triggered when the Toomre $Q$ value (*36*) $Q \equiv c_s\Omega/(\pi G\Sigma)$, falls below unity, possibly leading to very efficient angular momentum transport by spiral density waves and shocks (*37*). However, the onset of the GI requires a high surface density which, if present, likely occurs only in the very early phases of PPDs [e.g., refs. (*38, 39*)]. Spiral density waves and shocks can also be generated from a massive outer companion [e.g., ref. (*40*)] to which can angular momentum can efficiently transported when the density jump across the shock approaches order unity (*41*); however, this scenario is unlikely to be applicable to the solar nebula as Jupiter is unlikely to have accommodated the angular momentum of the entire solar nebula (which is expected to be more than 10 times more massive than Jupiter and extending well beyond Jupiter's orbit). Recent high-resolution disk surveys have reported a paucity of spiral patterns (*5, 6*), consistent with the general absence of GI in the bulk disk population. Another possibility is the Rossby wave instability (*42*), which is triggered in localized regions such as pressure extrema or disk gap edges. Although the Rossby wave instability tends to develop into large-scale vortices in the vicinity of such regions (*43, 44*), it is unlikely to affect angular momentum transport at the global scale.

To summarize, a range of pure hydrodynamic instabilities may operate in PPDs under certain thermodynamic conditions. While we do not fully understand whether they exist and, if so, their potential effects on PPDs, current studies tend to suggest that they are either not present for the bulk of the PPD lifetime or that they do not provide sufficient angular momentum transport to account for typical PPD accretion rates.

## C. Magnetohydrodynamic mechanisms
**1. Overview.** Magnetic fields have long been thought to play an important role in the gas dynamics of protoplanetary disks (*45*) and mechanisms involving magnetic fields are now thought to be primarily responsible for angular momentum transport in PPDs. As mentioned earlier, the magnetic field can contribute to radial and vertical transport of angular momentum via the $R\phi$ and $z\phi$ components of the Maxwell stress tensor, respectively. Even without knowledge of detailed disk microphysics, some important constraints can already be obtained relating accretion rates and magnetic field strength (*46, 47*).

More specifically, if accretion is mainly driven by radial transport through the $R\phi$ component of the Maxwell stress, we have $\dot{M}_{\mathrm{acc}} j = 2\pi R^2 L_z (\overline{B_R B_\phi}/4\pi)$, where the overbar again indicates time and spatial averages, and $L_z$ is the total thickness over which the stress is exerted [e.g., on the order of a few scale heights, $H \equiv c_s/\Omega_k$ (*48*), with $H\sim0.03$AU at 1AU for a Sun-like star]. $B_\phi$ is expected to be the dominant field component because any radial field is easily sheared to produce $B_\phi$. Ignoring $B_z$ and assuming that on average, $B_\phi$ is a factor $f > 1$ larger than $B_R$, we obtain [analogously to equation. (16) of ref. (*47*)]:

$$B_{\mathrm{mid},R\phi} \cong 0.72 \text{ G } (M/M_\odot)^{1/4} (\dot{M}_{\mathrm{acc}}/10^{-8} M_\odot y^{-1})^{1/2} (fH/L_z)^{1/2} (R/\mathrm{AU})^{-11/8} \qquad (2)$$

where $B_{\mathrm{mid}}$ is the midplane field. Taking $f \sim 50$ and $L_z \sim 6H$ (see Section 4), one finds the pre-factor becomes 2.0 G.

On the other hand, if accretion is mainly driven by vertical transport (via a magnetized disk wind emerging from the disk surface), then assuming symmetry about the disk midplane, one finds $\dot{M}_{\mathrm{acc}} j = 8\pi R^3 |\overline{B_z B_\phi}/4\pi|_{\mathrm{base}}$, where the subscript indicates the field strength at the base of the wind (usually several scale heights above and below the midplane). Similarly, assuming that on average, $B_\phi$ is a factor $f' > 1$ larger than $B_z$ at the wind base, we obtain [analogously to equation (7) of ref. (*47*)]:

$$B_{\mathrm{mid},z\phi} = mB_{\mathrm{base},z\phi} \cong m \ (0.065 \text{ G}) (M/M_\odot)^{1/4} (\dot{M}_{\mathrm{acc}}/10^{-8} M_\odot y^{-1})^{1/2} f'^{1/2} (R/\mathrm{AU})^{-5/4} \qquad (3)$$

where we further assume midplane field is some factor $m$ of the field at the wind base. Taking $f' \sim 10$ (see Theory Section C4), one finds the pre-factor becomes 0.21 G.

A few remarks are in order regarding the above relations. First, these results are fairly general and are independent of disk surface density. Radial transport only weakly depends on disk temperature (through the disk scale height). Second, given similar field strength, vertical transport is more efficient than radial transport by a factor of $R/H$, which is due to the large lever arm ($R$) for wind-driven accretion, whereas the $R\phi$ stress is exerted only over the disk thickness $L_z \sim$ a few $H$. Third, while (2) and (3) offer separate constraints on field strengths for the two mechanisms, (2) expresses field strength averaged over the bulk disk and (3) indicates surface field strength. Fourth, the two mechanisms can co-exist, such that (2) and (3) would reflect their individual contributions to accretion rates. Fifth, unlike previous expressions relating field strength to accretion rate (*10, 47*), equations (2) and (3) are not lower limits but instead direct estimates of the field with uncertainties reflected in the factors of $f$, $f'$, and $m$, respectively.

**2. Magnetic mechanisms.** The main physical mechanisms responsible for radial and vertical transport are the magnetorotational instability (MRI) (*49*) and magnetized disk winds (*50*), respectively. A basic understanding of these two mechanisms can be obtained from two physical processes.

The first is the decomposition of the Lorentz force per unit volume (*51*). In the ideal magnetohydrodynamic (MHD) limit (i.e., for well ionized gas; see below), it is given by:

$$\boldsymbol{F} = \frac{1}{c}\boldsymbol{J} \times \boldsymbol{B} = \frac{(\nabla \times \boldsymbol{B}) \times \boldsymbol{B}}{4\pi} = \boldsymbol{\kappa}\frac{B^2}{4\pi} - \nabla_\perp \frac{B^2}{8\pi} = \boldsymbol{F_T} + \boldsymbol{F_P} \qquad (4)$$

where $\boldsymbol{\kappa} \equiv \boldsymbol{b} \cdot \nabla \boldsymbol{b} = -\boldsymbol{R}_c/R_c^2$ is the curvature vector of the field with curvature radius $R_c$, $\boldsymbol{b}$ is a unit vector along the direction of $\boldsymbol{B}$, and $\nabla_\perp$ denotes the components of the gradient perpendicular to $\boldsymbol{B}$. The two terms on the right are the magnetic tension, $\boldsymbol{F_T}$, and magnetic pressure, $\boldsymbol{F_P}$, both of which are perpendicular to $\boldsymbol{B}$. Equation (4) shows that magnetic field lines tend to straighten (act against bending) and spatial variations in magnetic field strength exert a pressure that can expel gas towards weaker field regions.

The second physical process is that Keplerian shear in the disk constantly generates toroidal field from radial field. In the ideal MHD limit, the magnetic induction equation $\partial \boldsymbol{B}/\partial t = \nabla \times (\boldsymbol{v}_K \times \boldsymbol{B})$ (see below) for a Keplerian velocity profile, $\boldsymbol{v}_K$, yields $\partial B_\phi/\partial t = -(3/2)\Omega_K B_R$.

For the MRI, consider two fluid elements along a vertical field line that are located at different heights in the disk (Fig. 3A). If one element is slightly displaced radially inward, it will orbit slightly faster, while if the other is displaced radially outward, it will orbit more slowly. The radial field resulting from this displacement is then sheared to produce toroidal field. The curvature in this field builds up magnetic tension, attempting to bring the two elements back to their initial positions. The tension thus acts like a spring, with the tension growing as the two elements are increasingly separated. However, the torque resulting from the spring reduces/increases the angular momentum of the inner/outer element, making it spiral inward/outward. The process runs away, giving rise to the MRI. In ideal MHD, the growth of the MRI saturates by generating vigorous turbulence (*52*) that can transport angular momentum radially outward with typical values $\alpha \sim 10^{-2}$ in the absence of a net vertical field threading the disk (*53*). With a net vertical field, $\alpha$ at saturation increases with disk magnetization and can even reach order unity (*54*). Because the $R\phi$ Maxwell stress usually dominates over Reynolds stress by a factor of ~4-5, accretion under these conditions is mainly magnetically-controlled.

Magnetized disk winds require the disk to be threaded by a net vertical (poloidal) field. Because parent molecular clouds are magnetized (*55*), their fields may have been inherited by the disks as they are likely dragged in and bent radially outwards during cloud collapse (*2*). At the disk surface layer, shear in the disk constantly generates toroidal field from the radial component of the outward-bent poloidal field (but with opposite signs at the two disk sides), making the toroidal field increasingly stronger. This has two consequences (Fig. 3B). First, magnetic pressure gradually builds up, and eventually pushes the gas away from the disk, launching a wind. Note that this interpretation applies when the vertical field is relatively weak like that in PPDs (*11*), in contrast to the originally-proposed bead-on-wire centrifugally-driven wind that requires a strong vertical field (*50*). Second, the process that amplifies radial into toroidal field increasingly pinches the poloidal field lines in the disk against the direction of rotation. Magnetic tension associated with this pinched field exerts a torque on the disk that opposes disk rotation, thereby extracting disk angular momentum corresponding to the $z\phi$ component of the Maxwell stress. The property of the wind is primarily controlled by the strength of the poloidal field threading the disk, with stronger field extracting angular momentum faster.

**3. Complications from microphysics.** Considering more physical details leads to substantial complications. PPDs are distinguished from other astrophysical accretion disks (e.g., black holes) in that they are extremely poorly ionized, such that the coupling between gas and magnetic field is weak. This is because PPDs are generally too cold for collisional (thermal) ionization to occur unless temperatures exceed ~$10^3$ K (*56*). Instead, the bulk disk relies on nonthermal sources of ionization, particularly cosmic rays (*57*) stellar X-rays (*58*), and stellar ultraviolet radiation (*59*). The typical ionization fraction is of order $10^{-10}$-$10^{-14}$ in the midplane at several AU and boosted to $10^{-4}$-$10^{-5}$ in the disk atmosphere at a wide range of radii. Weakly-ionized gas experiences non-ideal MHD effects such that the magnetic field is no longer frozen into the fluid. There are three non-ideal MHD effects: ohmic resistivity, the Hall effect, and ambipolar diffusion (AD) (*46, 60*). Physically, they correspond to electron-neutral collisions, electron-ion drift, and ion-neutral drift, respectively, which allow field lines to drift relative to the gas and/or to dissipate in different ways. They are reflected in the nonideal MHD induction equation (*46, 60*), which determines how magnetic fields evolve:

$$\frac{\partial \boldsymbol{B}}{\partial t} = \nabla \times (\boldsymbol{v} \times \boldsymbol{B}) - \frac{4\pi}{c}\nabla \times (\eta_O \boldsymbol{J} + \eta_H \boldsymbol{J} \times \boldsymbol{b} + \eta_A \boldsymbol{J}_\perp) \quad (5)$$

where $\boldsymbol{v}$ is fluid velocity (i.e., velocity of the neutrals), $\boldsymbol{J} = (c/4\pi)\nabla \times \boldsymbol{B}$ is the current density, $c$ is the speed of light, and the subscript $\perp$ represents the component perpendicular to $\boldsymbol{b}$. The three terms in the right parentheses are the ohmic, Hall and AD terms, with corresponding magnetic diffusivities $\eta_O, \eta_H$ and $\eta_A$. The strength of non-ideal MHD effects is characterized by the three Elsasser numbers, $\Lambda_{O,H,A} \equiv v_A^2/(\eta_{O,H,A}\Omega_K)$, where $v_A = B/\sqrt{4\pi\rho}$ is the Alfvén speed. These effects are considered strong when the corresponding Elsasser numbers are less than or equal to order unity. In most cases, it can be shown that $\eta_O \propto x_e^{-1}$, $\eta_H \propto x_e^{-1}(B/\rho)$, $\eta_A \propto x_e^{-1}(B/\rho)^2$, where $x_e$ is the ionization (i.e., electron) fraction. This specifies the relative importance of the three effects: ohmic resistivity dominates in the densest regions (midplane of inner disk up to ~10 AU, where $\Lambda_O$ is well below unity), AD dominates in the most tenuous regions [the surface layer of inner disk and the entire outer disk, where $\Lambda_A$ is found to be of order unity (*60*)], and the Hall-dominated region lies in between.

Significant progress has been made in the past decade in understanding how individual non-ideal MHD effects affect the local gas dynamics, particularly on the MRI, as well as the global dynamics when all these effects are combined with realistic ionization chemistry. A detailed description is again beyond the scope of this review. Here we highlight some main results.

Ohmic resistivity is the best-studied. It suppresses the MRI when $\Lambda_O$ falls below order unity (*61*). When only ohmic resistivity is considered, it creates a "dead zone" in the midplane region of the inner disk (*62, 63*). AD can suppress the MRI for a strong field and damp the MRI for a weak field. The threshold field strength decreases with increasing AD (*64*). When ohmic resistivity and AD are combined, it is found that the MRI in the inner disk (~1-10 AU) is entirely suppressed in the disk vertical column (*65*). As a result, angular momentum transport requires the presence of a large-scale poloidal field threading the disk, which launches the magnetized disk wind. The wind is launched from the disk surface layer, where gas and magnetic fields are better coupled due to the higher level of ionization from stellar far-ultraviolet radiation (*65, 66*). Note that because wind-driven accretion is highly efficient in angular momentum transport, only a weak net vertical field is needed to drive accretion at desired rate of $\sim 10^{-8} M_\odot$ y$^{-1}$, corresponding to a plasma $\beta$ (defined as the ratio of the gas pressure to the magnetic pressure) of the order $10^5$ in standard solar nebular models. This picture likely also applies to the outer disk (beyond ~10 AU), which is entirely dominated by AD except that the outer disk can be weakly MRI turbulent (*67, 68*)

Equation (5) shows that unlike in the ohmic or AD-dominated regimes, in the Hall-dominated regime ($\eta_H$ significant compared to $\eta_O$ and $\eta_A$), the field evolution depends on the polarity of the vertical disk field [this is also in contrast to the geodynamo (*69*)]. When the net vertical field is aligned with disk rotation (i.e., $\boldsymbol{B} \cdot \boldsymbol{\Omega} > 0$), it results in the Hall-shear instability (HSI), which strongly amplifies the horizontal (both radial and azimuthal) fields (*70-72*). In this context, the magnetic field has an effective drift (Hall drift) speed, $v_H$, in the direction of $-\boldsymbol{J}$ as can be inferred from equation (5) (this is essentially the electron-ion drift speed). To understand the HSI, consider an initially vertical magnetic field in the disk perturbed in the $\phi$-direction (Fig. 3C). The Hall effect preserves the amplitude of the perturbed field and the Hall drift makes this pattern of magnetic perturbation rotate counterclockwise, producing a radial field component. The shear flow then generates toroidal field from the radial field that reinforces the perturbation, leading to instability. With both radial and toroidal fields amplified, the HSI yields a strong $R\phi$ Maxwell stress that leads to efficient radial transport of angular momentum. Unlike the MRI, the HSI itself does not generate turbulence and associated Reynolds stresses although it can enhance turbulence if the MRI were active (*73-75*). In PPDs, the Hall effect is always accompanied by ohmic resistivity or ambipolar diffusion, with the disk remaining largely laminar and launching disk winds. The wind also advects horizontal field out of the disk, which balances field growth and hence determines the saturation level of the HSI. As a result, contributions to angular momentum transport from $\mathcal{T}_{R\phi}$ and $\mathcal{T}_{z\phi}$ are typically found to be comparable. In the case of opposite polarity ($\boldsymbol{B} \cdot \boldsymbol{\Omega} < 0$), the Hall drift leads to clockwise rotation and subsequent shear will generate toroidal fields that reduce the original perturbation. As a result, the horizontal field is reduced towards zero (*70, 73*), thus contributing negligibly to angular momentum transport, although the configuration can be subject to a non-axisymmetric instability that produces bursts of weak turbulence (*76*).

Theoretically, we anticipate that the initial conditions for star formation should not distinguish between the aligned and anti-aligned cases. However, it has been shown that the Hall effect also plays an important role during disk formation stage, resulting in different disk initial sizes depending on the polarity (*77, 78*). Thus, disks with different magnetic polarities are likely born differently, although it is unclear whether such bimodality is present in observed disk populations. As such, current theoretical studies treat the two cases as equally probable.

Recently, global disk simulations that incorporate all three non-ideal MHD effects have been achieved (*79, 80*). Fig. 4 shows simulation snapshots under standard PPD parameters for both field-aligned and anti-aligned cases from ref. (*79*). It was found that in the aligned case, the horizontal field in the disk inner region (~1-10 AU) is amplified so much by the HSI that it does not change sign until reaching the surface at one side of the disk. Towards the outer region where the Hall effect (and hence the HSI) weakens, the change in sign occurs at the midplane. Transport of angular momentum by laminar Maxwell stress (radially) and by disk winds (vertically) make similar contributions. For the anti-aligned case, the field configuration also shows a high level of asymmetry. The toroidal field is again the dominant field component, only changing sign at the very uppermost layer on one side of the disk. Without the HSI, the toroidal field is much weaker than for the aligned case. The bulk disk is weakly turbulent, although it only causes minor spatial and temporal fluctuations in the total field strength without affecting the overall field configuration [these fluctuations can be absorbed to the $f$ or $f'$ and $m$ factors in equations (2) and (3)]. In both the aligned anti-aligned cases, the wind plays a major (if not overwhelming) role in driving disk accretion.

**4. Relating paleomagnetic measurements to astrophysical constraints.** As discussed below, individual meteoritic sample (e.g., chondrule or bulk matrix sample) provides measurements of the paleomagnetic field strength. Accompanying information includes an estimate of the age and location (usually the midplane region at some radius) of the meteorite magnetic record, as well as the duration over which the magnetic record has been averaged (see next section). Constraining solar nebula conditions requires that paleomagnetic measurements be understood in the context of a model of the nebular magnetic field. This can be achieved either by directly comparing with the best available PPD simulations or by utilizing the more general relations (2) and (3). Direct comparison with simulations has the advantage of having access to information about the entire radial and vertical profiles of magnetic fields (as in Fig. 4). However, the physical parameters of the primordial solar nebula are poorly known. On the other hand, equations (2) and (3) are the most general but require proper understanding of the physical mechanisms that drive disk angular momentum transport and depend on physical parameters of $f$, $L_z$, $f'$ and $m$.

The best understanding may be to combine both approaches, where we start from the framework of equations (2) and (3) and quantify the uncertainties in the predicted value of $B$ using the best-available simulations. Below we discuss the situation in the inner disk (i.e., $R \lesssim 10$ AU). Based on recent simulations incorporating all non-ideal MHD effects, it is found that when the net vertical field is aligned with the disk rotation, the midplane field is dominated by $B_\phi$, which is amplified to order $10^2$ times $B_z$ while $B_R$ and $B_z$ are comparable in strength. At the wind-launching region, $B_\phi$ is about a factor of 10 times $B_z$. In the anti-aligned case, $B_\phi \sim 10 B_z$ throughout the disk (from the midplane to the wind base). Since wind-driven accretion dominates, we may use equation (3) and take $f' \sim 10$ for a rough estimate of the wind base toroidal field for both cases. Additionally, we may take $m = 10$ and 1 for aligned and anti-aligned cases, respectively. For the aligned case, given that radial transport also makes significant contributions, we may also use equation (2), taking $f \sim 50$, $L_z \sim 6H$ which yields a field strength similar to the estimate from equation (3) at 1 AU. Simulations suggest that varying the fiducial parameters (e.g., surface density, magnetization thermodynamics, and ionization and diffusivity models) modestly changes these ratios. These variations can lead to changes in $f$, $L_z$ $f'$, and/or $m$ in equations (2) and (3) by a factor of a few, such that we estimate an overall factor of up to ~3 uncertainty in the field strength from these considerations.

Three additional issues are worth mentioning. First, there may be modest localized spatial variations in the field. In particular, $B_\phi$ must change sign and go through zero when traversing vertically across the disk, at which location the disk field largely consists of just the vertical component. This means that the field values predicted from (2) and (3) would be overestimates within a narrow range of disk heights. This may be more of an issue in the outer disk where the current sheet likely lies around the midplane (see also Fig. 4). The net vertical magnetic field may also be subject to localized radial variations on scales of several disk scale heights superimposed on the general radial trend (*81, 82*), although the variation may be more modest for total field strength. The total field strength may also vary in response to local variations such as condensation fronts (*82*) and, probably more importantly, from planet formation (although the latter has not been well quantified). Although the uncertainty associated with such spatial variations can be considered as different realizations of the previously mentioned uncertainties and hence can be readily be absorbed into the uncertainty associated with the factors $f$, $f'$ and $m$, here we conservatively increase the uncertainty in field strength from the aforementioned factor of ~3 to now a factor of ~5.

Second, the background field may experience temporal variations which could be recorded by quickly-cooled meteoritic materials as a range of paleointensities that are individually snapshots of the field. Although existing global simulations have been run only up to a few thousand years, it appears that at least in the inner disk, no significant changes (e.g., in the direction of $B_\phi$) occur other than some secular changes in total field strength from the slow, long-term evolution of magnetic flux. Therefore, the time-averaged (over at least > 1000 y) and instantaneous field strengths at a given location in the disk are expected to be similar.

Third, if the meteoritic object (e.g., chondrule) is itself spinning rapidly relative to the field variations, it will record the projection of the background field on the body's spin axis, which will on average reduce the apparent paleointensity by a factor of 2 for a population of bodies with randomly oriented but non-tumbling spin axes (*83*). The uncertainty associated with this effect can be reduced to a negligible level if a sufficient number of paleomagnetic measurements are measured and averaged from chondrules formed within a restricted location and time window.

Additional uncertainties arise from the disk accretion rate. Astronomical observations can only measure instantaneous accretion rates onto the protostar which are not necessarily the same as the accretion rate at a certain radius. Furthermore, the measured stellar accretion rates are typically uncertain by a factor of 2-3. Individual disks can also differ significantly from each other and there is an overall trend that accretion rate decreases with age, reaching up to $10^{-6}$ $M_\odot y^{-1}$ at the youngest ages (in the so-called Class 0 phase in the first ~$10^5$ y), down to well below the standard value of $10^{-8}$ $M_\odot y^{-1}$. In particular, disks with young ages ($\lesssim 1$ Ma) likely experience FU Orionis-like (typically Class 0) or EXOr-like (typically Class I/II) outbursts, which are interpreted as episodic boosts of the accretion rate by a factor of up to a few hundred or up to a few tens, respectively (*84*). Our ignorance about nebular accretion rates limits the use of paleointensities to constrain nebular physics. A median value of ~$10^{-8}$ $M_\odot y^{-1}$ for a 1 $M_\odot$ star is a reasonable estimate for the bulk lifetime of the solar nebula (i.e., Class II phase), modulo uncertainties by up to a factor of a few. Given that the field strength depends on the square root of the accretion rate [see equations (2) and (3)], this uncertainty is ~$3^{1/2}$ unless we are considering the accretion rates at the very early phases of the solar nebula or near the end of disk lifetime. Combining all of the above uncertainties (i.e., associated with the factors of $f$, $L_z$, $f'$ and $m$, spatial variations of the field, and accretion rate), we estimate that at most times and locations in the disk, the total uncertainty on the predicted field as function of distance is less than or equal to an order of magnitude.

These uncertainties apply to predictions of the nebular field strength from theory and simulations to compare with measured paleointensities. Conversely, as discussed below, paleointensity measurements can be used to constrain accretion rates or the formation location of the meteorites, with uncertainty factors again propagating according to equations (2) and (3).

Multiple constraints can help reduce uncertainties. For example, multiple paleomagnetic measurements from samples originating from different heliocentric distances and different ages can be considered collectively in the context of the theory. In the meantime, astronomical observations of young PPDs are also on the verge of providing useful constraints to disk magnetic fields, as we discuss next.

**ASTRONOMICAL OBSERVATIONS**
As mentioned earlier, disk magnetic fields are likely inherited from the star formation process. The existence of magnetic fields in molecular clouds and protostellar cores is well established [see (*55*) for a review]. This was achieved mainly through two techniques: observations of dust polarization and of the Zeeman effect.

In the interstellar medium (ISM), any irregular-shaped dust grain in an anisotropic radiation field can be spun up around its short axis. The spinning grain can then acquire a magnetic moment along this axis due to surface charges and/or the Barnett effect (a spinning body with unpaired electrons can be spontaneously acquire a magnetization proportional to rotation rate) (*85*). In the presence of a magnetic field, its spin axis will then precess around the magnetic field and eventually align with the field due to dissipation [e.g., ref. (*86*)]. Once aligned, dust extinction of initially unpolarized background light after passing through the dusty region would be strongest for polarization oriented perpendicularly to the sky-projected field orientation, $\boldsymbol{B}_{POS}$. This means that background starlight should exhibit linear polarization along $\boldsymbol{B}_{POS}$. On the other hand, thermal emission from the dust (mostly at wavelengths comparable to or longer than the dust grain size) is primarily linearly polarized along the long axis that is perpendicular to $\boldsymbol{B}_{POS}$. Observations of both kinds of emission are extensively used to study magnetic fields in the ISM and star-forming regions and have revealed a link between magnetic field orientations, gas densities, and kinematic structures on a variety of scales [see refs. (*87-89*)].

The Zeeman effect describes the splitting of spectral lines into multiple components of slightly different frequencies in the presence of a magnetic field. The amount of splitting is proportional to *B*. However, for the ISM field strength, the splitting is usually too small to be spectrally resolved, such that its effect is mainly exhibited as circular polarization whose polarization degree is proportional to the line-of-sight component of the field, $B_{LOS}$. In the ISM, the Zeeman effect has been detected in atomic H, OH, and CN lines, as well as in masers, all in sub-mm to radio wavelengths. These measurements have measured field strengths over a wide range of densities from ~10 $\mu$G from diffuse clouds to up to a few mG in molecular cores [e.g., ref. (*55*)].

Primarily, efforts to measure magnetic fields in disks have focused on measuring polarized dust continuum emission, which is expected to largely probe the disk midplane region (for optically thin dust) and also to sense the surface of edge-on systems. This approach is challenging partly because disks are small (about $10^{-3}$ the size of molecular clouds and $10^{-2}$ the size of prestellar cores) and hence resolution and sensitivity requirements are highly demanding. Early attempts either yielded only upper limits (*90, 91*) or positive detections whose inferred field configuration is somewhat puzzling under the assumption of magnetic alignment (*92-94*).

With respect to theory, because of grain growth and the fact that gas in PPDs is much denser than that in the molecular cloud, it is likely that dust spin becomes collisionally damped before it can precess around the magnetic field towards alignment (*95, 96*). Additionally, several mechanisms have been proposed to produce sub-mm continuum polarization without involving magnetism. In particular, it has been realized that thermal emission from (sub-mm sized) dust can be scattered by other dust (called self-scattering) to yield linearly polarized radiation (*97, 98*). The polarization pattern and polarization fraction depend on disk inclination, and can be used to constrain dust properties.

The situation has been revolutionized with the advent of the Atacama Large Millimeter/submillimeter Array (ALMA). With orders of magnitude improvement in resolution and sensitivity, it has enabled spatially-resolved measurements of linear polarization pattern for nearby PPDs. More than a dozen disks have been observed so far, showing polarized intensity and pattern in the majority of cases that is consistent with self-scattering (*9, 99, 100*). However, the situation is more controversial for some observations at longer wavelength of 3 mm (*101, 102*), pointing to other mechanisms such as radiative alignment (*96*) and aerodynamic alignment (*103*), although these interpretations are also not without problems (*104*). In addition, polarization in two young and edge-on systems appears to be consistent with multiple origins and magnetic dust alignment remains compatible with the data (*105*). Overall, there is a paradigm shift in the disk community that dust continuum polarization has a commonly nonmagnetic origin, but the detailed mechanisms are still under active research.

Recently, ALMA has opened up the possibility of measuring circular polarization and it has become possible to constrain the line-of-sight field strength via the Zeeman effect. Modeling suggests that using the CN line [which is likely the most sensitive tracer of the surface layer and outer region (*106*)], a positive detection requires an area-averaged mean line-of-sight field strength of the line-emitting region to be at least 1-10 mG given the current sensitivity of ALMA (*107*). So far, several observations are being conducted, although no positive detection has been reported (*108*). The fact that the disk magnetic fields are likely smooth (as the MRI is largely suppressed) minimizes cancellations along the line of sight. In the meantime, from our earlier discussions, we anticipate younger disks with higher accretion rates (exceed $10^{-8}$ $M_\odot \text{y}^{-1}$) to be promising candidates.

In addition to the above, molecular line emission (i.e., rotational transitions) can become linearly polarized either parallel or perpendicular to the magnetic field due to the Goldreich-Kylafis (GK) effect (*109*). The GK effect results from the magnetic sublevels of the excited state deviating from local thermodynamic equilibrium due to an anisotropic radiation field, which can be external or internal in origin (e.g., from an anisotropic velocity gradient). It can be used to map out the orientation (but not intensity) of the plane-of-sky magnetic field, but there is a 90° directional ambiguity. The GK effect has been detected in evolved stars (*110*) and star-forming regions including molecular outflows (*111, 112*). Forward modeling suggested that the effect can potentially be detected in protoplanetary disks (*113*). Very recently, marginal detections towards two disks by ALMA have been reported (*114*) although the data are insufficient to infer field morphology.

# PALEOMAGNETISM
## A. Introduction
### 1. Natural remanent magnetization (NRM)

A unique opportunity to study nebular magnetism is offered by our solar system, for which we have planetary materials that recorded ancient magnetic fields. Their record is in the form of NRM, a macroscopic vector quantity that reflects the semi-permanent alignment of electron spins. NRM in planetary materials can in principle persist for far longer than the age of the solar system, long after the decay of the magnetizing field (*115*).

Paleomagnetic records from solar system materials are obtained from laboratory analyses of meteorites and in situ spacecraft magnetometry of small bodies. Paleomagnetic studies are complementary with respect to astronomical observations for characterizing nebular magnetism. Meteorite studies probe the conditions in our own well-studied early solar system while astronomical studies can characterize a diversity of young stellar objects. Furthermore, as mentioned above, paleomagnetic measurements likely constrain the field in the midplane where planetesimal formation and radial-azimuthal accretion occurs, while astronomical observations using the Zeeman effect would mainly provide area-averaged line-of-sight field components of the line-emitting region, likely tracing the disk surface.

Meteorite studies have four principle advantages over astronomical observations. First, paleomagnetic measurements constrain fields at extremely high spatial resolutions (down to the 0.1 mm or smaller scale of chondrules and inclusions). Second, unlike the model-dependent field strengths for dust polarization studies, they enable direct and highly accurate measurements of field paleointensities (in principle to 10% or better). Third, paleointensities typically have very highly accurate and precise ages (with uncertainties as little as 0.1 Ma) due to the ability to date the samples using radiometric methods. Fourth, because meteoritic materials have acquired their NRMs over timescales ranging from hours to millions of years or longer depending on the sample, they can probe the time-averaged field over a wide range of timescales along with the instantaneous field measured astronomically.

On the other hand, there are several disadvantages of paleomagnetic studies relative to astronomical observations. First, because the original orientations of meteoritic materials in the nebula are unknown, only the field paleointensity and not its paleodirection can be inferred. Second, the locations at which planetary materials acquired their NRMs may be difficult to constrain. Third, meteoritic materials may have been derived from only a small fraction of the volume of the solar nebula whose local conditions might not reflect the mean state of the overall disk.

Small bodies and meteorites with paleomagnetic records dating back to the early solar system are thought to mainly contain one of four forms of NRM when they cool, crystallize, or accrete in the presence of a steady ambient paleomagnetic field, $B_{paleo}$. Igneous materials (e.g., chondrules) and thermally-metamorphosed materials acquire a thermoremanent magnetization (TRM) during cooling below the Curie temperature of their constituent ferromagnetic minerals (*116*). Materials in which ferromagnetic crystals grow and/or experience recrystallization at low temperatures acquire a crystallization remanent magnetization (CRM) (*116*). Both TRM and CRM have been identified in a diversity of meteorites. Compression-decompression cycles due to impacts may have also imparted shock remanent magnetization (SRM), although such magnetization has not yet been unambiguously identified in natural samples (*117*). Finally, it has recently been proposed that meteoritic materials could acquire an accretional detrital remanent magnetization (ADRM) as resulting from compass-needle alignment of their constituent grains during gentle accretion (*118*). Although ADRM has not yet been recognized in meteorites, its terrestrial counterpart, known as detrital remanent magnetization, is ubiquitously observed in clastic sediments on Earth.

For typical geologic and meteoritic materials, NRM is approximately proportional to the intensity of the ancient magnetizing field for the typically weak ($\lesssim 10$ G) fields expected in the nebula (*116*):

$$M_{\text{NRM}} = \chi B_{\text{paleo}} \tag{6}$$

where $M_{\text{NRM}}$ denotes NRM and $\chi$ is the remanence susceptibility, a constant that depends on both the nature of ferromagnetic grains in the sample and the form of magnetization.

To obtain $B_{\text{paleo}}$ from laboratory paleomagnetic measurements, $M_{\text{NRM}}$ is measured after removing any partial magnetization overprints by degaussing and/or heating the sample in zero field. Then, because $\chi$ is not known a priori, the sample is remagnetized in a known lab field, $B_{\text{lab}}$, followed by measurement of the resulting magnetization, $M_{\text{lab}}$. Assuming $\chi$ remains unchanged by this process, then the two resulting equations can be divided to obtain $M_{\text{NRM}}/M_{\text{lab}} = B_{\text{paleo}}/B_{\text{lab}}$ and solved for $B_{\text{paleo}}$ (*116*). In principle, this requires that $M_{\text{lab}}$ have the same form of magnetization (i.e., TRM, CRM, SRM or ADRM) as $M_{\text{NRM}}$ because each form of magnetization will have a different value of $\chi$. However, reproducing the magnetization process for ADRM and most forms of SRM and CRM is currently essentially impossible in the laboratory given the unknown conditions in which these forms of magnetization were acquired by meteorites and the difficulty of recreating such conditions even if they were known. This is far more straightforward for obtaining paleointensities from natural TRMs, which involves heating and cooling the samples in the laboratory. Even so, even such heating experiments are challenging since they can cause the samples to undergo thermochemical alteration that leads to changes in $\chi$. To alleviate the requirement that the laboratory remanence be produced by the same process that produced the NRM, a sample can instead be given an analog magnetization like saturation isothermal remanent magnetization (IRM) (exposure to a strong field at room temperature) or anhysteretic remanent magnetization (ARM) (exposure to an alternating field with superimposed DC bias field at room temperature); if the ratio of the $\chi$ value for IRM to that of TRM or for ARM to TRM is independently known [e.g., through calibration experiments on analog samples ref. (*119*)], a paleointensity estimate can be obtained with typical uncertainties of between a factor of ~2 and up to an order of magnitude (*120-122*). Because CRMs usually have lower intensities than TRMs acquired in the same field (at least for ferromagnetic minerals that did not form as transformation products of pre-existing ferromagnetic minerals), only a minimum paleointensity can usually be inferred from CRM-bearing meteorites (*123*).

## 2. Rock magnetism
Ferromagnetic minerals contain unpaired electron spins whose orientations are mutually coupled by quantum mechanical exchange interactions. As such, they can form crystals with nonzero magnetic moments that can record an ambient magnetic field in the form of NRM. The most common ferromagnetic minerals in meteorites used for nebular magnetic field studies are the metals kamacite ($\alpha$-Fe$_{1-x}$Ni$_x$ for $x < \sim 0.05$), martensite ($\alpha_2$-Fe$_{1-x}$Ni$_x$ for $\sim 0.05 < x < \sim 0.25$), and awaruite (Fe$_{1-x}$Ni$_x$ for $\sim 0.67 < x < \sim 0.75$), the iron oxide magnetite (Fe$_3$O$_4$) and iron sulfides like monoclinic pyrrhotite (Fe$_{1-x}$S$_x$ for $x < \sim 0.13$) (*124-126*). Kamacite and martensite typically dominate the remanence of unmetamorphosed ordinary chondrites and most basaltic achondrites, whereas magnetite and sulfides carry much of the remanence in carbonaceous chondrites as well as in some achondrites like angrites.

A critical requirement for successful paleomagnetic studies is the identification of samples with high-fidelity magnetic recording properties. Most importantly, this means that the samples should (i) have been capable of acquiring an NRM proportional to the paleofield as described by equation (1), (ii) be able to retain this magnetization over at least 4.5 billion years, and (iii) be able to be cleaned of overprinting remanence by laboratory thermal or alternating field (AF) demagnetization. It has long been recognized that materials with a significant fraction of crystals with sizes in the single domain range (whose electron spins are uniformly aligned within the grain) are optimal magnetic recorders in this way (*116*). However, for some ferromagnetic minerals like kamacite, the single domain size range can be vanishingly narrow for grains with equant shapes (*127*). As a result, FeNi-bearing meteoritic materials are instead commonly dominated by larger grains like those in the single vortex state (for which the spins typically take the form a uniformly oriented core surrounded by a spiral structure). Nevertheless, very recent advances in micromagnetic modeling have now established that most single vortex grains can carry exceptionally high-stability NRM that can be stable over the history of the solar system (*128*).

Several different kinds of materials have recently been identified to likely carry high-fidelity records of nebular fields. In particular, a small fraction of chondrules (typically 10% in ordinary chondrites, and even rarer in carbonaceous chondrites) contain sub-micrometer-sized crystals of nearly pure-Fe kamacite embedded in forsterite called "dusty olivines" (Fig. 5). This metal is thought to have formed during a reduction process prior to accretion. The metal's fine grain size means that much of it is in the single vortex size range (*129-131*). Furthermore, its very low Ni abundance (<2 wt. %) guarantees that the metal grains did not undergo subsolidus phase changes that can occur during slow cooling on the parent body long after accretion and which may alter the paleomagnetic record (*132, 133*). As such, dusty olivine-bearing chondrules that remained below the 780°C Curie point of kamacite following formation are almost certain to contain a total TRM record of nebular magnetism.

Another class of high-fidelity ferromagnetic minerals in meteorites are magnetite and pyrrhotite grains that typically formed during post-accretional aqueous alteration on the parent body. These grains may contain a CRM (for meteorites that remained at low temperature during and after metasomatism like CM chondrites) or a TRM for heated meteorites (like metamorphosed CM chondrites). Like dusty olivine metal, magnetite and pyrrhotite in meteorites commonly form grains in the single domain and single vortex size range, but they have the additional advantage of usually being much more resistant than FeNi to thermochemical alteration during laboratory heating (see above). Another limitation is that because magnetite and pyrrhotite typically usually formed after accretion of the parent body, it can be challenging to distinguish whether their NRMs were acquired in in the nebular field or that of a dynamo generated within the meteorite parent body (*134*).

## 3. Remagnetization processes

A central challenge faced by paleomagnetic studies is to establish that the NRM in a given sample formed in the early solar system in the presence of a nebular magnetic field or instead is a more recent overprint. Although atmospheric passage typically only thermally remagnetizes the outer <0.3 cm of stony meteorites, many meteorites are later essentially completely remagnetized by collector's hand magnets and weathering on Earth's surface (*125, 135*). Furthermore, even those meteorites that avoided remagnetization on Earth may have already been remagnetized by thermal metamorphism, aqueous alteration, and/or phase changes and heating from impacts on their parent bodies following the dissipation of the nebula.

There are several ways that the age of the NRM can be established. Radiometric ages can constrain the timing of thermal and crystallization events. Petrographic and mineralogical analyses can constrain relative formation ages of ferromagnetic minerals based on their textures, compositions, and overprinting relationships with other materials in a meteorite. For example, sulfide veins that transect from a chondrule's interior into the surrounding matrix [as observed CV chondrites (*136*)] must postdate accretion. Furthermore, measurements of mutually-oriented subsamples of a given meteorite can provide direct relative age constraints on the time of NRM acquisition (*137*). For example, the demonstration that the NRM direction in the fusion crust and adjacent baked interior zone is distinct from that of the deeper interior would indicate that the meteorite has not been completely remagnetized as part of a single remagnetization event after landing on Earth (*125*). Furthermore, because the nebular field is not expected to exceed ~5 G in most locations [e.g., equations (2) and (3)], a demonstration that the ratio of NRM to saturation IRM is < 10% indicates that the sample has likely escaped exposure to a hand magnet (*122, 125, 138*). Finally, and most importantly, the demonstration that relative magnetization directions of individual chondrules or other clasts and inclusions are collectively random (while each individual object has magnetization that is uniformly oriented within it) would indicate that their NRM predates assembly of the meteorite (*83*). For chondrules, which are expect to have accreted orientations onto their parent body with orientations that are randomly oriented relative to the paleofield direction (assuming no ADRM forms), the observations of scattered magnetization directions can be taken as powerful evidence that their NRMs formed prior to accretion.

**B. Meteorite measurements**
**1. Overview.** Although the paleomagnetism of meteorites has been studied since the 1950s (*139*), the nature and meaning of meteorite NRM have only begun to be understood in detail in the last decade. In particular, the identification of ferromagnetic minerals thought to have formed in the early solar system along with tests of NRM stability and age (see above) have recently enabled the identification of early solar system magnetic field records in four chondrite parent bodies (LL, CM, CV, and CR groups) and two achondrite parent bodies (angrites and NWA 7325). We now review the paleointensity records from these meteorites. We only discuss samples for which rock magnetic analyses, radiometric ages, and constraints on the remagnetization history have established that the NRM is ancient (e.g., no younger than 6 Ma after the formation of the solar system).

**2. Ages of NRMs.** The ages of the NRM records in these meteorites are precisely constrained by a variety of short-lived and long-lived radionuclide systems (Table S1). Because we are interested in the ages of the field records relative to the formation of the solar system rather than their absolute ages, we have calculated ages relative to CAIs, which are the oldest known solids and which are thought to have formed within $10^5$ y of the collapse of the molecular cloud (*140*). The meteorite materials used for dating are chondrules, secondary alteration minerals (magnetite and carbonate), and bulk igneous rocks (as well as separates from these materials). The crystallization, thermal metamorphism and aqueous alteration events experienced by these materials have been dated using both the long-lived U-Pb system, which provides absolute ages, and several short-lived systems (Al-Mg, I-Xe, Mn-Cr, and Hf-W), which provide ages relative to a standard whose absolute U-Pb age is independently known (*141*). The use of relative chronometers relies on the assumption that the initial abundance of the short-lived parent nuclide was homogenous in the solar system where the standard and the sample of interest formed. However, relative chronometers have the advantage over U-Pb ages in typically having higher precision due to the much shorter half-lives of their parent nuclides. A second advantage of the Al-Mg and Hf-W systems in particular is that CAIs themselves are commonly used as a standard, meaning that uncertainties associated with the absolute age of CAIs do not contribute to the uncertainty budget for these systems. This is relevant because estimating the uncertainty of an age relative to CAIs should not only include the nominal measurement uncertainties of any CAI age, but also take into account the consideration that there are two different ages that have been obtained for CAIs which differ from each other by >3 standard deviations: 4567.30 ± 0.16 Ma ago from ref. (*142*) and 4567.94 ± 0.21 Ma ago from ref. (*143*) (with the caveat that only the former age has been described in a detailed refereed publication).

Meteoritic materials acquired their NRMs over widely varying timescales. Therefore, their paleomagnetic records represent average magnetic fields over these time scalesß. Chondrules are thought to have typically cooled from the Curie point to ~0°C over timescales of ~1-$10^3$ h (*144-146*) such that they will acquire a near-instantaneous TRM record of the ambient field. By comparison, aqueous alteration of chondrites on their parent bodies is thought to have a duration of >1 y (*147*) and up to several Ma (*148*), such that post-accretional CRMs should provide a time-averaged record the field that could extend over a significant fraction of the nebula's lifetime.

**3. Locations where NRMs were acquired.** There are broad constraints on the formation locations of meteoritic materials that enable interpretation of their magnetic records in the context of nebular processes. With respect to their vertical positions in the disk, drag forces on 0.1 mm-sized grains in the nebula are expected to lead them to settle to within a disk scale height of <~0.04 AU of the midplane at 1 AU from the Sun within ~$10^4$ y (*149*) consistent with observational evidence of settling and the largely laminar environment expected from theoretical considerations (see above) (*150*). As such, it is expected (although not required) that most meteorites and their constituent chondrules and refractory inclusions should have recorded magnetic fields near the midplane. With respect to their radial positions in the disk, observations of meteoroid trajectories indicate that nearly all parent bodies were immediately derived from the asteroid belt (2-3 AU) (*151*). However, this does not require that the meteorites were in the present-day asteroid belt during their formation. In particular, it has recently been realized that known meteorites are derived from two groupings ("carbonaceous" and "noncarbonaceous") with distinct O, Cr, Ti, Mo, and W isotopic compositions that are thought to sample two nebular reservoirs [e.g., ref. (*152*)]. It has been proposed that these two reservoirs were located within and beyond proto-Jupiter's orbit, respectively, where they had become isolated when proto-Jupiter opened up a gap in the disk after reaching its isolation mass. In particular, an analysis of the abundance of CAIs and refractory elements in meteorites has predicted that noncarbonaceous and carbonaceous parent bodies accreted at ~2-3 and 3-4 AU (*153*), whereas the similar oxygen isotopic compositions of ordinary and carbonaceous chondrites suggests formation of the latter within <7 AU (*154*).

**4. Paleointensity results.** We now summarize existing paleointensity constraints on the nebular field (Table S2). Measurements of chondrules from the Semarkona LL chondrite (*10*) indicate the existence of a near-instantaneous field of intensity 0.54 ± 0.21 G at ~1-3 AU (*153*) at 2.03 ± 0.81 Ma after the formation of CAIs. If chondrules formed by nebular shocks associated with planetary bow shocks, they are expected to record a field intensity similar to that of the ambient unshocked nebula (*155*). Because the LL chondrules in Semarkona are magnetized in random directions, they must have been magnetized prior to accretion and therefore recorded the nebular field (see Paleomagnetism section A3)

Paleomagnetic studies of seven CM chondrites (*134*) indicate that they were magnetized by a field of >0.06 G at ~3-7 AU (*153, 154*) at 2.90 ± 0.39 Ma after CAI formation (*134*). The NRMs within the CM chondrites are unidirectional and carried by the magnetite-bearing aqueously altered matrix, requiring that were magnetized after accretion. They should have recorded the time-averaged field, consistent with the difference between CM I-Xe and Mn-Cr ages (Table S2). The post-accretional nature of the NRM in CMs means that the origin of the magnetizing field is somewhat ambiguous: it may have been a product of nebular field or conceivably a more recent field that of a core dynamo within a partially differentiated CM parent body [e.g., ref (*137*)]. Thermal modeling indicates that convective core dynamos in mantled planetesimals should be delayed by at least 4-5 Ma after accretion (*156, 157*) due to thermal blanketing of the core by the silicate mantle, which was heated by the decay of the short-lived radionuclide $^{26}$Al. However, such a delay need not apply to dynamos powered by other mechanisms like impacts (*158*) or mantle precession (*159*). Assuming the thermal blanketing constraint is relevant to the CM parent body, the ~2 Ma age of CM chondrite NRM would indicate that they recorded the nebular magnetic field. The CM data therefore suggest that the nebular field in the outer solar system lasted until at least ~2.51 Ma after CAI-formation (Table S1).

Recent studies of four slightly younger meteorite groups found no evidence of a nebular field (Fig. 7, Tables S1-S2). Paleomagnetic studies of chondrules from two CR chondrites demonstrate that they too pass fusion crust and conglomerate tests (*160*). However, due to the relatively low fidelity of their magnetic recording properties, only an upper bound on the instantaneous field of 0.08 G at 3.68 ± 0.22 Ma after CAI-formation at ~3-7 AU (*153*) can be placed on the nebular field from CR chondrules. The absence of primary magnetization in the ungrouped achondrite NWA 7325 indicates that it cooled in a field of instantaneous intensity <0.034 G at ~1-3 AU (*161, 162*) at 5.24 ± 0.05 Ma after CAI formation (*124*). Second, the absence of stable high blocking temperature magnetization in the Kaba CV chondrite indicates that the time-averaged ambient field was ≲0.003 G at ~3-4 AU (*153*) at 4.08 ± 0.77 Ma after CAI formation (*163*). Thirdly, the absence of primary magnetization in three angrites indicate that they too cooled in the absence of a nebular field (instantaneous value <0.006 G) at ~1-3 AU (*164*) at a mean age of ~3.71 ± 0.23 Ma after CAI-formation (*165*).

**C. Spacecraft measurements.** Spacecraft have measured the magnetic field at the surface of 2 asteroids, during flybys of 5 other asteroids, and at the surface of a comet (*166-168*). Although remanent magnetic fields have been tentatively by two of the asteroid flybys, none of these investigations found unambiguous evidence of a remanent field. Moreover, several investigations placed stringent upper limits on the mean asteroid NRM that are at the lower range of the values measured for cm-sized samples of meteoritic materials ($<2\times10^{-6}$ emu g$^{-1}$). However, because these missions only constrained the magnetization averaged over spatial scales ranging from 10-20 cm [as inferred from recent Hayabusa2 measurements across the surface of asteroid (162173) Ryugu (*167*)] to $10^5$ km$^3$ [as inferred from asteroid (21) Lutetia (*169*)], the intensity of the finer-scale NRM in these asteroids may be well above these limits. Furthermore, with the possible exception of the comet measurements, the timing of the magnetization records in these asteroids is unknown because of the lack of radiometric and other ages for their constituents. Therefore, the paleointensity constraints inferred from meteorites are not inconsistent with lack of detections of remanent magnetic fields around asteroids. An unprecedented opportunity to synthesize magnetic field datasets from spacecraft and laboratory measurements will be provided by the return of Hayabusa2 and OSIRIS-REx samples to Earth in the next few years (*170*).

The magnetic field measurements measured by the Rosetta lander Philae at the Jupiter-family comet 67P Churyumov-Gerasimenko offer a unique constraint on nebular magnetism. Other than being the only near-field measurements at a comet, the 67P measurements are also distinguished by the fact they were taken as part of a near-surface > 1 km-long transect with 4 separate landings during which the magnetometer sensor reached within 5 cm of the surface (*168*). Philae constrained the magnetization of comet to $<5\times10^{-6}$ emu g$^{-1}$ for spatial scales of >10 cm (*172*). Other data from Rosetta suggest that 67P accreted by pebble-pile processes, which would indicate the presence of the nebula during its assembly(*171*). In this case, 67P could have acquired ADRM in a sufficiently strong nebular field. Given evidence that solids in 67P and other Jupiter-family comets have chondrite-like compositions and mineralogies (*173, 174*), and assuming sucn an ADRM has been preserved since the comet formed, the very low NRM intensity would constrain the local nebular field to be <0.03 G (*172*). Given the expected formation locations of Jupiter-family comets like 67P, this would apply to somewhere in the region 15-45 AU.

## DISCUSSION
### A. Support for a central role of magnetism in stellar accretion?
The paleointensity constraints on the nebular field of 0.54 ± 0.21 G at 2.03 ± 0.81 Ma after CAI formation from LL chondrites and of >0.06 G at 2.90 ± 0.39 Ma after CAI formation from CM chondrites are both consistent with the field predicted by equations (3) and (4) for accretion rates of ~$10^{-8}$ $M_\odot$ year$^{-1}$ for a disk whose rotation vector and magnetic field are aligned (Section Theory Section B4) (Fig. 8). The expected field for the case of anti-aligned polarity for this accretion rate would be about an order of magnitude lower) (Fig. 8). The anti-aligned polarity would be consistent with the data if the disk accretion rate were on the order of ~$10^{-7}$ $M_\odot$ y$^{-1}$ or higher. This is not impossible if the disk evolution is far from steady-state, but otherwise if the bulk accretion rate is a non-increasing function of time as observations suggest (*7, 175*), the mass of the solar nebula would have to be unusually massive, at least reaching a substantial fraction of a solar mass.

Furthermore, the dependence of the field with distance as constrained by the LL chondrules, and CM bulk chondrites, and Philae measurements at 67P is broadly consisted with the predicted decrease in field with distance from the Sun ) (Fig. 8). In particular, the lower limit from CM chondrites together with the estimated field strength from LL chondrules are broadly consistent with a temporally constant accretion rate. However, this tentative inference requires more paleomagnetic data for confirmation.

### B. Distinguish between chondrule formation mechanisms?
Nebular paleomagnetic records also constrain how the first solids formed. In particular, chondrules are 0.1-1 mm-diameter igneous inclusions found in nearly all primitive, undifferentiated meteorites. Their ubiquity in the meteorite record suggests that they may have constituted a significant fraction of accreted mass in early-forming planetesimals, although this may reflect bias in the formation mechanisms and/or locations of chondrites. The formation mechanism of chondrules remains vigorously debated, with nebular shocks and planetesimal collisions being the subjects of most recent research.

Two observable features of chondrule NRMs can provide information about chondrule formation. First, the existence of primary, unidirectional NRM blocked across a range of temperatures implies that the orientation of the chondrule rotation axis and the background magnetic field remained stable over the time of NRM acquisition, which likely ranged between hours to less than ~10 days (*176*). The presence of such magnetization in Semarkona chondrules has been used to argue for a chondrule number density of less than $4 \times 10^{-2}$ cm$^{-3}$ during chondrule formation and background PPD temperatures of <230°C, which appears inconsistent with the current sheet and short circuit instability hypotheses for chondrule formation (*176, 177*). This result further implies that magnetic fields in the solar nebula during Semarkona chondrule formation were stable over hours to ~10-day timescales.

Second, the paleointensities derived from NRM components can be compared to those predicted for different chondrule formation mechanisms. Among these, the X-wind model posits the Sun as the energy source that led to chondrule melting, while the impact model invokes the energy derived from collisions between planetesimal bodies. Shock waves in the nebular gas, generated at the large scale by gravitational instabilities or at the local scale by planetary bow shock, have also been hypothesized to lead to chondrule formation. Predictions of strong magnetic fields ≥0.8 G from the X-wind model of chondrule formation appear inconsistent with values recovered from Semarkona and CM chondrules (*10*). Models of planetesimal impacts and bow shocks suggest that chondrules formed by these processes should record background nebular field intensities, while chondrules formed by large-scale nebular shocks might record a field amplified by a factor of up to 30 above the background (*155, 177*). This implies that both planetesimal collisions and bow shocks are simultaneously consistent with magnetically-mediated nebular transport [see equations (3) and (4)] and LL- and CM-derived paleointensities. Meanwhile, large-scale shocks may also be consistent with magnetic transport and the meteorite measurements if the field that magnetized CM meteorites is well significantly above the minimum paleointensities inferred from those meteorites (i.e., well above 0.06 G).

**C. Lifetime of the nebula?**
Because the sustenance of magnetic fields requires the existence of a conducting medium as a necessary (although not sufficient) condition, the dispersal time of the nebula may be similar to the time when strong magnetic fields like those expected for the nebula (see predictions in Figs. 7, 8) disappeared as inferred from the absence of paleomagnetism in meteorites younger than a certain age (*165*) (Fig. 7, 8 and Tables S1-S2). For example, the interplanetary magnetic field at 10 Ma after CAI-formation is estimated to have only been ~2 mG at 1 AU (*178*), which is more than order of magnitude below the weakest nebular fields predicted for an accretion rate of $10^{-9}$ $M_\odot$ $y^{-1}$. The minimum lifetime of the nebular field in the inner and outer solar system is constrained by the NRM in LL chondrules and CM bulk chondrites, which have minimum ages of 1.22 and 2.51 Ma after CAI-formation, respectively (95% confidence lower limits). These provide a minimum constraint on the lifetime of the nebular field in each location. The most temporally precise constraint on the maximum lifetime of the nebular field comes from the angrites (*165*). Using equations (2) and (3), their <0.006 G paleointensity constraint indicates that by 3.71 ± 0.23 Ma after CAI-formation, accretion rates dropped in the inner solar system (<3 AU) to <$10^{-12}$ $M_\odot$ $y^{-1}$ and <$10^{-10}$ $M_\odot$ $y^{-1}$ assuming the disk rotation and poloidal fields were aligned and anti-aligned, respectively. We favor the former upper limit since the LL chondrule data support the aligned case for our solar system. These results are broadly consistent with the <0.003 G constraint from the CV chondrite Kaba, which indicates a low accretion rate in the outer solar system (~3-7 AU) (*153, 154*) at 4.08 ± 0.77 Ma after CAI-formation. However, because the Kaba paleointensities are recorded by a CRM, this upper limit is uncertain (see above).

The <$10^{-12}$ $M_\odot$ $y^{-1}$ accretion rate inferred from angrites is below the slowest observed accretion rates for protoplanetary disks (*179-181*). We therefore place a 95% confidence upper limit on the lifetime of the nebula in the inner solar system of 3.94 Ma after CAI-formation (*165*) (Fig. 7), calculated as the sum of the weighted U-Pb age for two angrites and the 95% confidence limit on the angrite and CAI U-Pb ages published in the refereed literature combined in quadrature (Table S1). Likewise, we estimate a maximum lifetime for the nebula in the outer solar system of 4.89 Ma (Fig. 7), calculated as the sum of the Mn-Cr age for CV chondrules plus the 95% confidence limits on the Mn-Cr ages, the U-Pb age of the D'Orbigny standard and the U-Pb age of CAIs added in quadrature (Table S1). Overall, the results indicate that the nebula had locally dispersed sometime between 1.22 and 3.94 Ma after CAI-formation in the inner solar system, and between 2.51 and 4.89 Ma in the outer solar system, consistent with recent estimates of the lifetime of the nebular dust disk (*182*).

This result has implications for how the gas giants Jupiter and Saturn formed. There are two main formation models for giant planets. In the disk instability model, a sufficiently massive disk subject to the gravitational instability and rapid cooling fragments into clumps within <1000 y which then cool to become giant planets (*183*). By comparison, core accretion is typically a multistage process beginning with the protracted growth of a ~10-20 $M_\oplus$ followed by rapid runaway gas accretion that in total typically requires ≳ 1 Ma (*183-185*). The viability of the core accretion model requires that the disk lifetime must therefore exceed ~1 Ma, consistent with astronomical observations of young stellar objects (*3*). The minimum 1.22 Ma lifetime of the nebula established by paleomagnetic studies therefore permits both the disk instability and core accretion models for gas giant formation in our solar system.

**D. Constraining formation distances of meteorites?** It is conceivable that instead of using paleomagnetic studies to constrain the nebular field intensity, the unknown formation locations of meteorites could be constrained using paleomagnetic measurements if the radial dependence of the field intensity were independently known. In particular, if it is assumed that accretion is being driven by magnetic fields at an accretion rate of ~$10^{-8}$ $M_\odot$ y$^{-1}$, then the paleointensity measured from a given meteorite or body from an unknown location could potentially constrain the distance from the Sun at which it acquired its magnetic record. In support of this possibility, existing data, although sparse, are broadly consistent with the predicted radial dependence of the nebular fields (Fig. 8). There are numerous uncertainties with such an approach, beginning with the requirement that magnetic fields be driving accretion and also including assumptions about angular momentum transport (from $\mathcal{T}_{R\phi}$ versus $\mathcal{T}_{z\phi}$), the assumed accretion rate, and possibility of local field heterogeneities. As implied by the discussed above, the uncertainty on the predicted field as a function of the formation distance from the Sun has an uncertainty of at least an order of magnitude (*81*).

In this context, two recent studies of the ungrouped C2 chondrites Tagish Lake and WIS 91600 measured weak to null paleointensities (<0.0015 and 0.044 ± 0.028 G, respectively) despite the inference that they formed during the lifetime of the solar nebula (i.e., at <3-4 Ma after CAI-formation). Using the above logic, these studies proposed that these meteorites formed at the current location of Saturn or beyond. This is broadly consistent with predictions from Fig. 8 assuming a minimum accretion rate of $10^{-9}$ $M_\odot$ y$^{-1}$, the weaker midplane fields predicted for $R\phi$ stresses, and that the disk rotation and poloidal fields were aligned. If correct, this would provide evidence for large-scale radial mass transport sometime during the last few billion years and would support the proposal that these samples may be from comet-like or Kuiper belt-like parent bodies.

**OUTLOOK AND UNSOLVED QUESTIONS**

The study of protoplanetary disk magnetism is advancing rapidly. From a theoretical standpoint, the recent inclusion of all three diffusivities associated with non-ideal MHD effects in the state-of-the-art numerical simulations has fundamentally improved our understanding of the role of magnetic fields in driving angular momentum transport in disks. More work is necessary to extend these studies to three dimensions and to incorporate more realistic physics, especially for thermodynamics and ionization chemistry [e.g., refs. (*186, 187*)]. Moreover, the current paradigm requires the presence of a net vertical magnetic flux threading the disk, which directly determines the rate of angular momentum transport and the associated field strength. While such magnetic flux is likely inherited from the star formation process, it remains to be understood how it evolves in disks [e.g., ref. (*188*)], which is a more fundamental question relating to the long-term evolution of disks and disk magnetism. As we learn more in the near future, we anticipate that the overall interpretive framework outlined in this review will remain valid, while some of the uncertainties may be narrowed down.

On the observational side, the search for magnetic field signatures in disks continues. While polarized dust continuum emission has thus far been found to be largely of nonmagnetic origin, it still provides useful constraint on disk and dust properties. The marginal detection of the GK effect (*114*) opens up a new approach to potentially constrain the field morphology in disks. Anticipated near-future measurements of the GK effect for more sources and spectral lines will provide valuable information on our understanding of disk magnetic fields. Probably the most promising technique for magnetic studies is the prospect of directly inferring the line-of-sight field strength from measuring the circular polarization signature of CN (and possibly other molecular) lines due to Zeeman splitting from ALMA. Even though the extreme sensitivity requirement precludes spatially resolving the field, the information would still allow for directly comparing the field strength with instantaneous accretion rate to validate the theory.

The spatial and time dependence of the nebular field in our own solar system remain poorly understood. These gaps will be addressed with the continued paleomagnetic measurement of new meteorite groups. In particular, just 4 of the ~13 known chondrite groups and just 2 achondrite parent bodies have been studied from the perspective of constraining nebular magnetism. Measurements of a diversity meteorites from different locations in the disk would further constrain the time required for the nebular field to disperse within the inner ~7 AU of the solar system. Furthermore, within regard to chondrites, field records have been obtained from just chondrules and bulk matrix-rich samples. Future analyses of CAIs could provide field records from the beginning of solar system history. Such data could also be used to address the question of how CAIs formed (*189*): some formation locations (e.g., within <0.1 AU of the young Sun) predict they would record strong paleointensities while others (e.g., in the expanding envelope of a supernova) seem to predict near-zero field values.

In this regard, nebular field constraints from spacecraft measurements of small bodies like comet 67P are especially valuable because of the additional constraints on the parent body's formation location provided by the geologic context of in situ measurements. As such, future sample return missions, like those in progress to carbonaceous asteroids [e.g., ref. (*167*)], are even more exciting since they will combine the advantage of geologic context with the sensitivity and precision of laboratory measurements.

**H2: Supplementary Materials**
    **Table S1.** Radiometric age constraints on meteorite records of nebular magnetism
    **Table S2.** Paleointensity constraints on the solar nebula magnetic field from meteorites

**Acknowledgments.** We thank D. J. Stevenson and an anonymous reviewer for thoughtful reviews and M. E. Brown for editing the paper.  We also thank J. Biersteker, J.F.J. Bryson, T. Bosak, S. Desch, A. Johansen, N. Kita, T. Kleine, and O. Pravdivtseva, S. T. Stewart, F. Tissot, and J. Wisdom for helpful discussions. **Funding:** BPW thanks Thomas F. Peterson, Jr. for support. XNB acknowledges support from the Youth Thousand Talents Program. **Author contributions:** BPW and XNB contributed to all sections of the paper, with BPW organizing the overall manuscript and leading the sections on meteorite and asteroid magnetism and XNB leading the sections on theory and astronomical observations. RRF led the discussion section on chondrule formation and contributed to the meteorite observations. **Competing interests:** The authors declare no competing interests. **Data and materials availability:** All data are available in the published literature. Key dates and paleomagnetic constraints are compiled in the Tables S1-S2.

# Figures

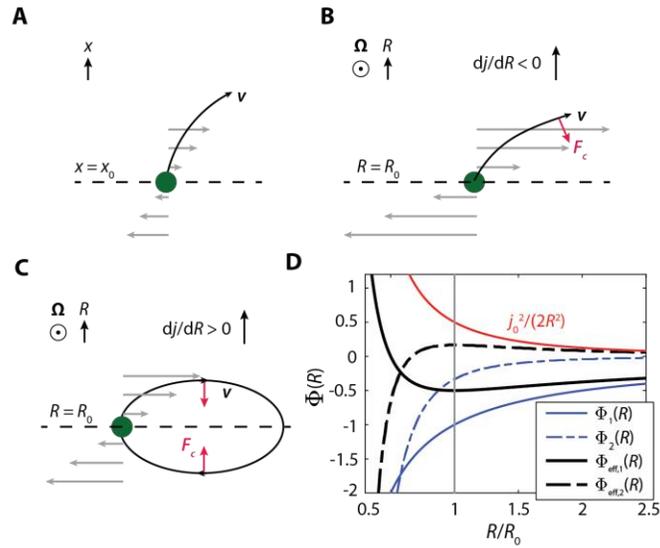

**Fig. 1.** Why accretion disks are linearly hydrodynamically stable. Consider three flows that contain a shear in fluid velocity, $v$ (grey). Arrow directions and magnitudes denote background flow direction and velocity. (A) In a non-rotating shear flow with even just a small shear, small displacements from of fluid parcels (green) from an initial position $x = x_0$ lead to unbounded motion (black arrow). (B) A rotating flow with angular velocity $\Omega(R)$ for radius $R$ in which the specific angular momentum $j = \Omega R^2$ decreases with $R$ (toward top of page) will form a relatively strong shear ($d\Omega/dR$). Although the Coriolis acceleration $\boldsymbol{F_c} = -2\boldsymbol{\Omega} \times \boldsymbol{v}$ (red arrow) tries to restore the motion back to the parcel's initial position at $R = R_0$, the competing effect of the strong shear leads to unbounded motion. This corresponds to potential $\Phi_2$ in (D). (C) In a rotating Keplerian sheer flow, $j(R)$ increases with $R$, leading to a reduced shear compared to that of case (B). As a result, the Coriolis acceleration (red arrow) confines the motion into epicyclic trajectories (black loop). This corresponds to potential $\Phi_1$ in (D). **(D)** Stability of a gas parcel on a circular orbit around $R = R_0$ (grey line) over two disk potentials $\Phi_1(R)$ and $\Phi_2(R)$, where $j_i(R) = \sqrt{R^3 d\Phi_i/dR}$ for $i = 1, 2$ increases with $R$ for $\Phi_1$ (as for a point mass potential) and decreases with $R$ for $\Phi_2$ (in this case, much more steeply than for a point mass potential for illustrative purposes). We have set $j_1(R_0) = j_2(R_0) \equiv j_0$ for simplicity. For a particle on a circular orbit, we have $j_0 = \Omega_0 R_0^2$. Due to angular momentum conservation, the parcel experiences as an effective potential that includes a centrifugal term $\Phi_{\text{eff},i} = \Phi_i(R) + j_0^2/(2R^2)$. It can be shown that $\Phi_{\text{eff},1}$ has a minimum at $R = R_0$, meaning that a circular orbit is stable to small perturbations, whereas $\Phi_{\text{eff},2}$ has a maximum at $R = R_0$, meaning that a circular orbit is unstable. Panels (A) and (C) after ref. (*17*).

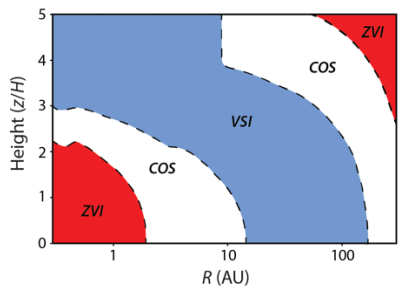

**Fig. 2. Regions in the disk where various hydrodynamic instabilities may operate.** Shown are the estimated locations where each of three hydrodynamic instabilities—the vertical shear instability (VSI) (blue), convective overstability (COS) (white) and zombie vortex instability (ZVI) (red)—is expected to dominate based on cooling time calculations for a specific disk model and under the assumption that magnetic fields are not present. Adopted from ref. (*15*) with permission.

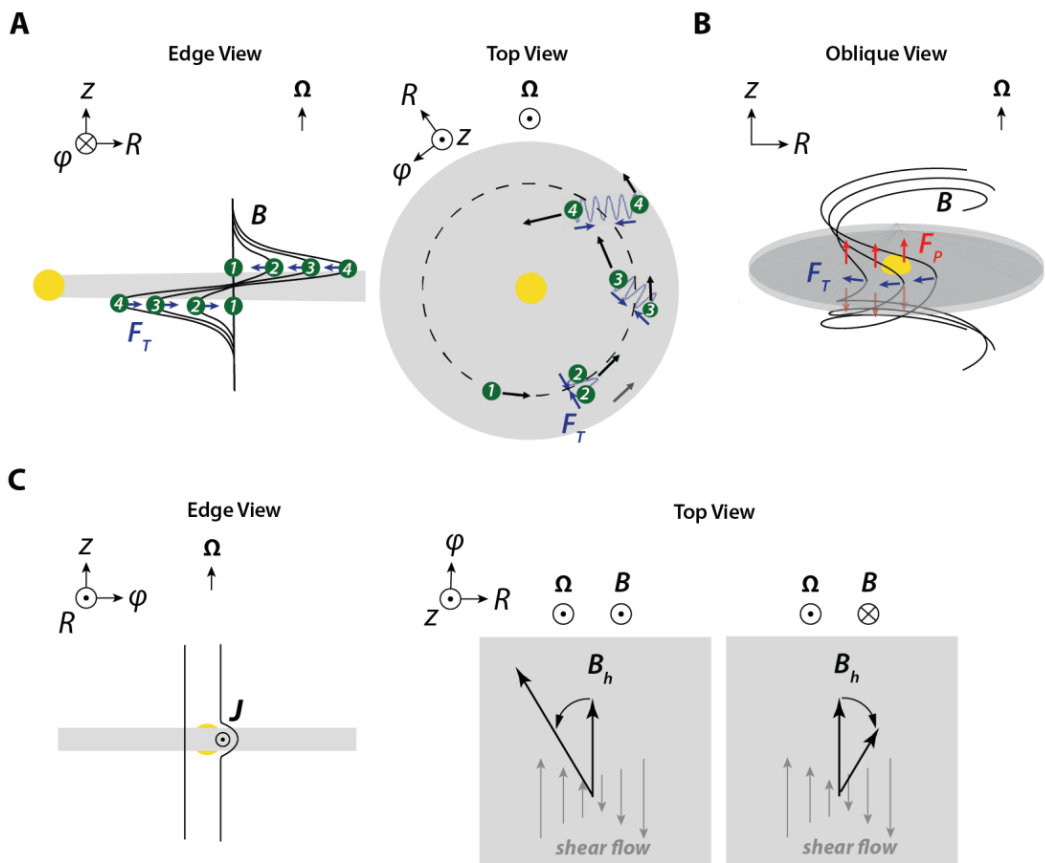

**Fig. 3. Angular momentum transport by magnetic fields in a protoplanetary disk**. (**A**) Transport by the magnetorotational instability. Shown is evolution of the instability at four different times (numbered). Left and right panels show edge-on and top views of disk. Direction of disk rotation, Ω, is shown at top left 1: Two gas parcels (green) are initially located at a narrow annulus in the disk (grey). 2: Slight displacement of one parcel inward and one outward, shown by black arrows, leads to a decrease in the outer parcel's Keplerian angular velocity and an increase

*Science Advances* Page 33 of 38

in the inner parcel's Keplerian angular velocity as they follow largely Keplerian motion. In doing so, the frozen-in nature of the magnetic field, $\boldsymbol{B}$ (black curves) leads to a spring-like tension force, $\boldsymbol{F_T}$ (blue arrows) that acts to attract the particles. 3-4: This force exerts a torque between the two parcels (exhibited as $R\phi$ Maxwell stress) that causes further drift of the inner parcel inward and the outer parcel outward. This process runs away, and eventually generating turbulence that also contributes to angular momentum transport via $R\phi$ Maxwell and Reynolds stresses. (**B**) Transport by the magnetized disk wind. 1: The disk is threaded by a poloidal magnetic field. 2. Keplerian shear of the radial component of the field produces a toroidal field in the disk. 3. The Lorentz force associated with the build-up (i.e., vertical gradient) of magnetic pressure, $\boldsymbol{F_P}$, launches a disk wind. In the meantime, the field is pinched through the disk against the direction of rotation, and the Lorentz force associated with this pinching, $\boldsymbol{F_T}$, extracts angular momentum from the disk. (**C**) Transport by the Hall shear instability. Left: starting with a vertical field embedded in the disk, a perturbation in the $\phi$ direction creates a radial current $\boldsymbol{J}$, which yields a Hall-drift velocity along -$\boldsymbol{J}$ direction. Right: This Hall drift makes the perturbed field rotate counterclockwise in the plane of the disk when the initial field is aligned with disk rotation (left subpanel), producing a radial field component (shown here is the horizontal field, $\boldsymbol{B_h}$). This radial field is then sheared to generate toroidal field that reinforces initial perturbation, leading to angular momentum transport via $R\phi$ Maxwell stress [e.g., analogous to right subpanel of (A)]. For an anti-aligned initial field (right subpanel), the combination of Hall drift and shear reduces initial perturbations. Right subpanel of (C) after refs. (*48, 72*).

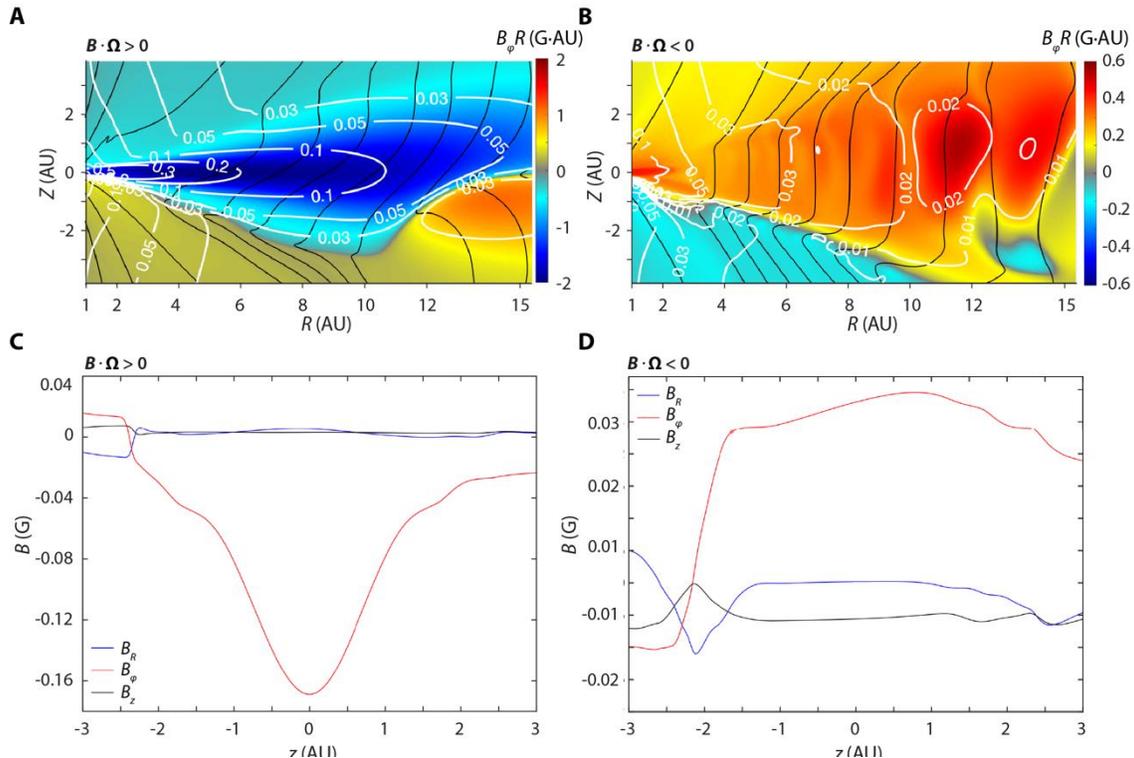

**Fig. 4. Numerical simulations of the magnetic field evolution in a protoplanetary disk.** (**A, B**) Magnetic field configurations from two-dimensional axisymmetric simulations incorporating all non-ideal MHD effects. (A) Vertical field aligned with disk rotation. (B) Vertical field anti-aligned with disk rotation. Shown in color is the scaled toroidal field strength $(B_\phi/\text{Gauss}) \cdot$

(*R*/AU). Black contours mark poloidal field lines and white contours mark the total field strength in Gauss. (**C, D**) Vertical profiles of the radial, azimuthal and vertical components of the magnetic field at 7 AU from the simulations shown in the top panels for vertical field aligned (C) and anti-aligned (D) with disk rotation. Simulations from ref. (*79*) after evolving about 1,200 and 1,800 y.

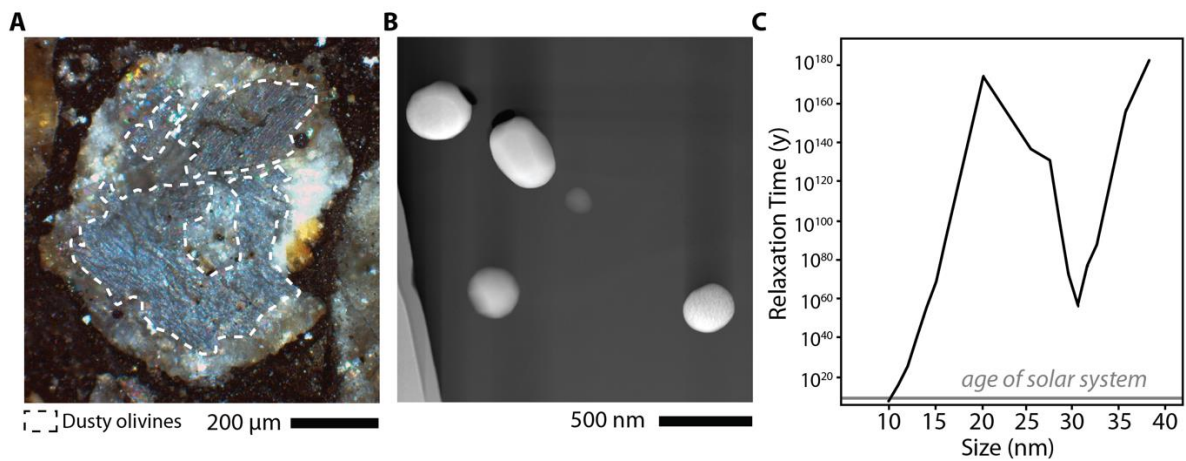

**Fig. 5. Dusty olivines as carriers of nebular magnetization.** (**A**) Reflected light optical photomicrograph of dusty olivine-bearing chondrule in the LL Semarkona meteorite. Light grey regions are mostly forsterite, whereas darker grey regions outlined by white dashed lines are rich in dusty metal. (**B**) Transmission electron microscopy image of four dusty olivine kamacite crystals (white ellipsoids) in forsterite (dark grey background). (**C**) Relaxation times at room temperature of kamacite cuboids with aspect ratios of 1.5 as calculated by micromagnetic modeling and compared to the age of the solar system (grey line). Grains larger than 25 nm in diameter are in the single vortex state while smaller grains are in the uniform (e.g., single domain or flower) state. Panels (A) and (B) after ref. (*10*) and panel (C) after ref. (*130*)

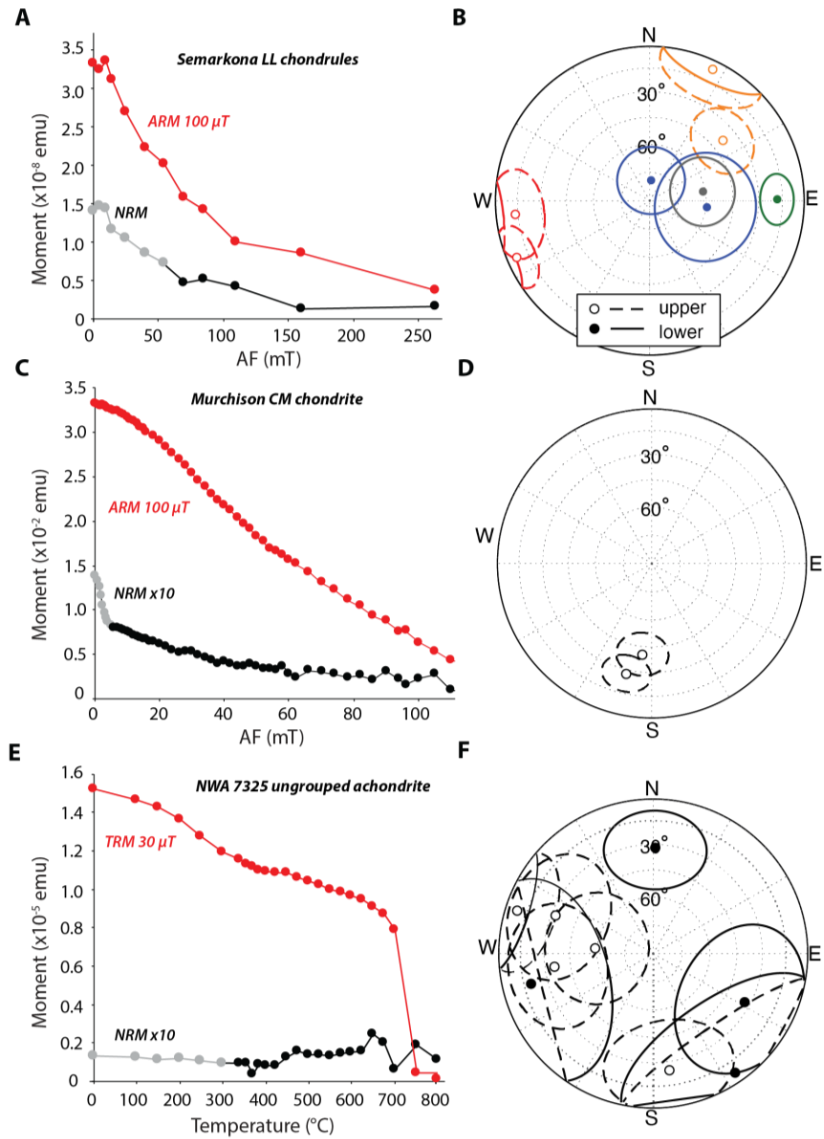

**Fig. 6. The nebular magnetic field as recorded by meteorites. (A, B)** Dusty olivine bearing chondrules from the LL chondrite Semarkona. Data from ref. (*10*). **(C, D)** Bulk samples of the CM2 chondrite Murchison. Data from ref. (*134*). **(E, F)** Bulk samples of the ungrouped achondrite NWA 7325. Data from ref. (*124*). Shown in (A, C, E) are AF demagnetization of NRM (with grey and black symbols denoting secondary and a high-coercivity components interpreted to be a record of nebular field conditions, respectively) and demagnetization of ARM (A, C) and TRM (E). Shown in (B, D, F) are equal area stereographic projections showing the directions (circles) and maximum angular deviation (MAD) values (ellipses; a measure of uncertainty) of the high-coercivity components in 5 mutually-oriented chondrules (B), 2 mutually-oriented bulk samples (D) and 9 mutually-oriented bulk subsamples (F). In (B, D, F) open symbols and dashed lines denote upper hemisphere projections and closed symbols and solid lines denote upper hemisphere projections. In (B), individual chondrules are denoted by distinct colors, with three chondrules yielding mutually two mutually-oriented subsamples each (orange, blue, and red). In (C and F), the NRM is multiplied by a factor of 10 for the purposes of visibility. The ARMs were acquired in a 1 G bias field with a 2,900 G AC field (A) and a 1 G bias field with a 1,000 G AC field (B),

while the TRM acquisition and thermal demagnetization were conducted by heating to 800°C in a field of 0.3 G in a controlled oxygen fugacity oven (set to 3 log units below the iron-wüstite buffer).

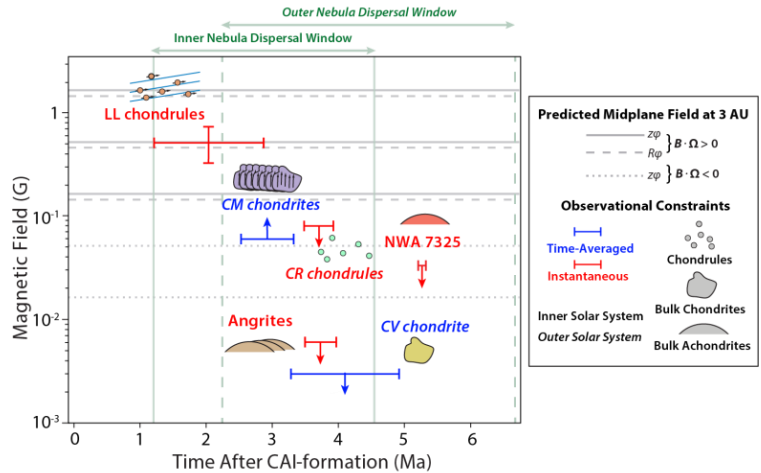

**Fig. 7. Paleomagnetic constraints on nebular field as a function of time**. Paleointensity of the nebular field as inferred from paleomagnetism of various meteorites. Downward- (upward-) pointing arrows indicate upper (lower) limits. Red and blue constraints denote paleointensity constraints on the instantaneous (red) and time-averaged (over >1 to $10^6$ y) (blue) fields. Magnetization in chondrules from the LL chondrite Semarkona (*10*), and in seven CM chondrites (*134*) indicates the nebular field persisted in the inner (1-3 AU) and outer (~3-7 AU) solar system until at least 1.22 and 2.51 Ma, respectively, after the formation of calcium aluminum-rich inclusions (CAIs) (vertical solid grey lines). The lack of detectable NRM in three volcanic angrites (*165*), the ungrouped achondrite NWA 7325 (*124*), and the high blocking temperature range of the CV chondrite Kaba (*163*) collectively indicate that the nebular field had dispersed by 3.94 Ma and 4.89 Ma after CAI formation in the inner and outer solar system, respectively (vertical dashed grey lines). This is consistent but not required by the lack of detectable NRM in CR chondrules (*190*). Horizontal lines show midplane magnetic field at 3 AU predicted from theory assuming magnetic stresses are driving accretion around a $1 M_\odot$ star. Solid and dashed lines denote field assuming the nebular field and sense of disk rotation are aligned, for which we assume contributions from both $R\phi$ [equation (2) with $f = 50$ and $L_z \sim 6H$] and $z\phi$ (dashed) [equation (3) with $f' = 10$ and taking $m = 10$] stresses, respectively. Dotted line denotes field assuming the nebular field and sense of disk rotation are anti-aligned, for which we assume contributions from just $z\phi$ (dashed) [equation (3) with $f' = 10$ and taking $m = 1$] stresses. For the aligned case, although the predicted field from the $R\phi$ stress is a factor of $R/H$ smaller than that for $z\phi$ stresses in the active layer, the predicted field at the midplane for the former is larger because $m = 10$. For each case, fields are estimated for three assumed different accretion rates: $10^{-9}$, $10^{-8}$, and $10^{-7} M_\odot \text{y}^{-1}$ (bottom, middle, and top lines). See Tables S1 and S2 for source data and details.

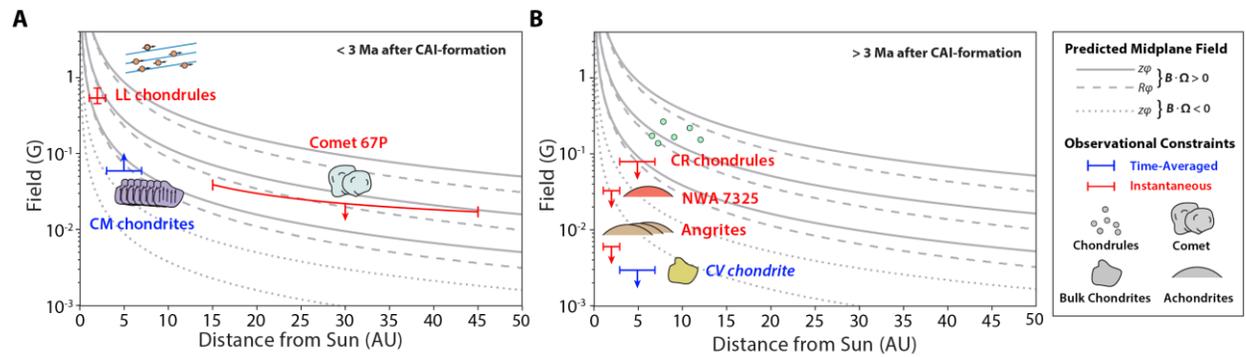

**Fig. 8. Paleomagnetic constraints on nebular field as a function of distance from the young Sun. (A)** Fields prior to 3 Ma after CAI-formation. **(B)** Fields after 3 Ma after CAI-formation. Red and blue constraints denote paleointensity constraints on the instantaneous (red) and time-averaged (over >1 to $10^6$ y) (blue) fields from chondrules in the LL chondrite Semarkona (*10, 155*), bulk CM chondrites (*134*), Philae measurements at comet 67P (*172*), CR chondrules (*190*), and bulk samples of NWA 7325 (*124*), angrites (*165*), and CV chondrites (*163*). Downward- (upward) pointing arrows indicate upper (lower) limits. Curves show midplane magnetic field predicted from theory assuming magnetic stresses are driving accretion around a $1 M_\odot$ star. Solid and dashed curves denote field assuming the nebular field and sense of disk rotation are aligned, for which we assume contributions from both $R\phi$ [equation (2) with $f = 50$ and $L_z \sim 6H$] and $z\phi$ (dashed) [equation (3) with $f' = 10$ and taking $m = 10$] stresses, respectively. Dotted curves denote field assuming the nebular field and sense of disk rotation are anti-aligned, for which we assume contributions from just the $z\phi$ (dashed) [equation (3) with $f' = 10$ and taking $m = 1$] stress. For each case, fields are estimated for three assumed different accretion rates: $10^{-9}$, $10^{-8}$, and $10^{-7} M_\odot \text{y}^{-1}$ (bottom, middle, and top curves). See Tables S1 and S2 for source data and details.

**Supplementary Materials**
Supplementary text
Tables S1 to S2